\documentclass[sigconf, nonacm]{acmart}

\usepackage{siunitx}
\usepackage{mhchem}
\usepackage{dblfloatfix}
\usepackage{hhline}
\usepackage{titlesec}
\titlespacing*{\subsubsection}{0pt}{\baselineskip}{\baselineskip}
\usepackage{balance}

\begin{document}
\title{Beyond Performance: Measuring the Environmental Impact of Analytical Databases}

\author{Michail Bachras}
\affiliation{%
  \institution{University of Toronto}
  \city{Toronto}
  \state{Canada}
}
\email{michalis.bachras@mail.utoronto.ca}

\author{Hans-Arno Jacobsen}
\affiliation{%
  \institution{University of Toronto}
  \city{Toronto}
  \country{Canada}
}
\email{jacobsen@eecg.toronto.edu}

\begin{abstract}
    The exponential growth of data is making query processing increasingly critical for modern computing infrastructure, yet the environmental impact of database operations remains poorly understood and largely overlooked. This paper presents ATLAS, a comprehensive methodology for measuring and quantifying the environmental footprint of analytical database systems, considering both operational impacts and manufacturing costs of hardware components. Through extensive empirical evaluation of four distinct database architectures (DuckDB, MonetDB, Hyper, and StarRocks), we uncover how fundamental architectural decisions affect environmental efficiency. Our findings reveal that environmental considerations in database operations are multifaceted, encompassing both immediate operational impacts and long-term sustainability implications. We demonstrate that architectural choices can significantly influence both power consumption and environmental sustainability, while deployment location emerges as a critical factor that can amplify or diminish these architectural advantages.
\end{abstract}

\maketitle

\pagestyle{plain}
\section{Introduction}

The exponential growth of digital data ~\cite{emirler2023world} has made efficient query processing more critical than ever before. At the heart of this challenge lies the analytical query processing - a complex orchestration of computational resources where CPUs run at peak frequencies, memory banks rapidly cycle data, and storage devices stream information~\cite{abadi2013design}. Yet beyond this visible choreography lies a hidden environmental impact that the database community has largely overlooked: each query execution triggers a cascade of both carbon emissions and water consumption through the power generation required to fuel these operations. This environmental footprint is not merely a byproduct of computation but it is intrinsically shaped by the fundamental architectural decisions we make in database system design, particularly significant given that data centers and ICT infrastructure account for a substantial portion of global energy consumption~\cite{jones2018stop,matthew2024global} and consume enormous amount of potable water~\cite{mytton2021data}.

The environmental impact of database operations manifests through two critical dimensions that demand our attention. First, carbon emissions arise not only from operational power consumption but also from the manufacturing of the hardware our systems require. Second, and perhaps more surprisingly, water consumption emerges as a crucial metric through the power generation process, from cooling thermal power plants to driving hydroelectric turbines. This water-energy nexus in database operations represents a critical consideration in database design: different architectural choices lead to distinct power consumption patterns, which in turn translate to varying environmental footprints.

What makes this consideration particularly compelling for the database community is how deeply it intertwines with our core research areas. Query optimization has traditionally aimed to minimize response time and resource usage. However, a new dimension has emerged: environmental efficiency. Incorporating environmental impact into database system design presents unique challenges that extend beyond traditional optimization metrics. 

A key aspect of this challenge lies in how power grids vary dramatically in their composition across regions, leading to significant variations in both carbon emissions and water consumption per unit of energy. The water footprint of power generation varies by orders of magnitude across different energy sources - from mere liters per megawatt-hour for wind power to thousands of liters for nuclear and hydroelectric generation. This geographical variability means that identical database operations can have drastically different environmental footprints based solely on deployment location. For instance, a database deployment in a region powered by hydroelectric energy will have a fundamentally different water consumption profile compared to one in an area relying on nuclear power, even when running identical workloads. These variations introduce a critical dimension to database system evaluation: we must understand how architectural choices interact with local power grid characteristics to determine overall environmental impact.

In this context, the database community stands at a critical crossroad where environmental consciousness must become integral to system design. The exponential growth in data volume~\cite{emirler2023world} makes this transformation urgent.  Yet achieving true environmental efficiency requires more than superficial changes - we cannot simply append "green" metrics to existing optimization frameworks. Instead, we must broaden how we architect analytical database systems by elevating environmental impact to an important design consideration. This evolution demands new methodologies for measuring environmental impact, optimization techniques that consider both computational and environmental efficiency, and system designs that can intelligently adapt to varying deployment contexts. 

Toward this vision, we present foundational work that advances our understanding of database systems' environmental impact. Our research introduces systematic approaches and novel insights for measuring, understanding, and quantifying the environmental impact of analytical database systems. Our contributions include:

\begin{enumerate}
    \item ATLAS (AnalyTicaL datAbase Sustainability), a comprehensive methodology for quantifying the environmental footprint of analytical database operations. Unlike existing approaches that focus solely on energy consumption, ATLAS captures both carbon emissions and water consumption through a novel framework that considers operational impact, hardware manufacturing costs, and geographical variations in power grid characteristics. 
    
    \item EcoQuery, the first sustainability-focused benchmark for analytical database systems. Built upon industry-standard benchmarks, like TPC-H, EcoQuery extends traditional performance metrics with environmental impact measurements, providing a standardized way to evaluate and compare the environmental efficiency of different database architectures. 
    
    \item An extensive empirical evaluation across four distinct database architectures - DuckDB, MonetDB, Hyper, and StarRocks - reveals previously unknown relationships between architectural design choices and environmental impact. We demonstrate how different query processing strategies, storage models, and resource management policies affect both carbon emissions and water consumption. 
    
    \item A novel analysis of how geographical location influences database environmental impact. By examining deployments across multiple regions with varying power grid compositions, we quantify how the same database operations can have dramatically different environmental footprints depending on deployment location.

\end{enumerate}

The remainder of the paper is organized as follows. Section 2 reviews related work in database energy efficiency, carbon emissions, and environmental impact measurement. Section 3 presents the necessary background while Section 4 details the environmental metrics that we use throughout this paper. Section 5 introduces the ATLAS framework and EcoQuery benchmark, which are used to evaluate the environmental efficiency of database architectures across various hardware configurations and geographical locations. Section 6 presents the experimental evaluation and analysis of the environmental impact of the examined database systems. Finally, Section 7 concludes the paper with a summary of findings and future research directions.

\section{Related Work}

The shift toward sustainability and environmental impact awareness in the ICT sector has resulted in an expanding corpus of study into exploring and optimizing the energy efficiency of database systems. Harizopoulos et al.~\cite{stavros2009energy}  propose some optimizations to reduce energy waste in database systems and point out directions for potential energy-efficiency optimizations. Tsirogiannis et al. \cite{tsirogiannis2010analyzing} analyzed the energy efficiency of a database server, characterizing power consumption as a function of workload characteristics, such as query operators and
parallelism. They show that different query operators have significant power consumption variations, emphasizing the non-linear relationship between CPU power and the number of cores used. Similarly, Psaroudakis et al.~\cite{psaroudakis2014dynamic} analyze the energy consumption of individual database operators, introducing metrics like Throughput per Watt and Energy Delay Product (EDP) to evaluate energy efficiency. 

Other works~\cite{willis2009towards,mahajan2019improving,karyakin2017analysis}, investigate techniques to improve the energy efficiency of different components of the underlying database server, such as CPU, DRAM and disks, while Liu et al~\cite{liu2021understanding} explore the energy efficiency of power-constrained hardware when running database operations. Kissinger et al.~\cite{kissinger2018adaptive} examine the energy consumption of an in-memory database and propose an adaptive energy control architecture that optimizes energy efficiency and performance of the underlying DBMS. Studies have explored the energy usage and efficiency in analytical~\cite{meza2009tracking,crotty2021case} and transactional~\cite{poess2008energy} workloads while other research works~\cite{xu2010exploring,lang2011rethinking} focus on investigating the tradeoff between energy consumption and query performance.

Aside from energy efficiency research, several recent approaches have emerged to quantify and reduce the environmental impact of software systems using metrics such as carbon emissions. Tools like CodeCarbon~\cite{electricitymap}, cloud-based carbon estimation frameworks~\cite{dodge2022measuring}, and architectural modeling systems~\cite{gupta2022act} demonstrate this shift beyond pure energy conservation toward thorough carbon accounting. CodeCarbon enables real-time estimation of code snippet emissions, while Dodge et al.~\cite{dodge2022measuring} focus on AI workloads using location-based and time-specific data. Gupta et al.~\cite{gupta2021chasing} investigate both operational energy consumption and hardware manufacturing, highlighting how the growing overall carbon footprint persists despite efficiency innovations, with significant emissions attributed to manufacturing and infrastructure. This comprehensive approach underscores the need for including both operational and embodied carbon emissions in environmental assessments. Building on this, ACT~\cite{gupta2022act} was developed to quantify and optimize the end-to-end carbon footprint of computing systems. Additional frameworks~\cite{patterson2021carbon},~\cite{henderson2020towards}, and \cite{uddin2015evaluating} further explore methods for tracking and optimizing the carbon footprint of AI workloads. 

However, the aforementioned literature focus on software systems other than database systems. Only one study~\cite{kannan2023towards} specifically examines the carbon footprint of database systems, focusing on NoSQL systems. Their work distinguishes carbon emission optimization from traditional performance metrics and investigates how DRAM configurations affect carbon emissions in Graph500 benchmark executions, considering both operational and embodied carbon emissions.

To the best of our knowledge, our work is the first to present a novel yet practical methodology for evaluating the energy efficiency and environmental impact of analytical database systems. Our approach seeks to clarify the influence of varying database architectures on the environmental footprint, considering not only the energy consumption during execution but also the hardware requirements of each database system. More specifically, we investigate multiple scenarios for executing database operations under high load conditions to identify the environmental surcharge and reveal potential optimization opportunities. Furthermore, our analysis assesses the environmental efficiency of each database, taking into account both operational factors and the underlying hardware infrastructure.

\section{Background}
\subsection{Database systems}
This study focuses on analytical database systems specifically designed for complex query processing and data analysis workloads. Our selection criteria emphasizes systems with distinct architectural approaches to analytical processing, particularly those that have influenced modern database design through documented technical innovations. Traditional relational database management systems like MySQL and PostgreSQL, while widely deployed, are not included as they primarily target online transaction processing (OLTP) workloads and implement row-oriented architectures that are not optimized for analytical operations.
\subsubsection{\textbf{DuckDB}}\hspace{-9pt}
DuckDB~\cite{raasveldt2019duckdb} is an in-process SQL OLAP (Online Analytical Processing) database system. It implements a columnar storage structure and uses vectorized query execution, processing data in blocks rather than row-by-row. As an embedded database, it operates within the host application process, eliminating client-server communication overhead. Version 0.9.1 is used in this study.

\subsubsection{\textbf{Hyper}}\hspace{-9pt}
Hyper~\cite{kemper2011hyper,funke2014hyper} is a main memory database system that combines transactional and analytical processing capabilities. Its architecture utilizes in-memory processing and columnar storage format. The system implements hybrid transactional/analytical processing (HTAP) functionality for concurrent OLTP and OLAP operations. Hyper employs parallel execution strategies and includes cost-based optimization with dynamic execution plans. This study utilizes the Hyper API provided by Tableau~\cite{hyperdocs}, as Hyper itself is not open-source.

\subsubsection{\textbf{MonetDB}}\hspace{-9pt}
MonetDB~\cite{boncz2008breaking,nes2012monetdb} is an open-source columnar database management system.  Its storage model is based on Binary Association Tables (BATs), where each BAT consists of an object identifier (OID) column and an attribute value column. The system implements an operator-at-a-time execution model, executing each operator to completion before initiating subsequent operators. Version 11.49 is used in this study.

\subsubsection{\textbf{StarRocks}}\hspace{-9pt}
StarRocks~\cite{starrocks} is an open-source MPP (Massively Parallel Processing) analytical database. Its architecture partitions data across compute nodes and divides database operations into parallel computing units. The system implements separate storage and compute layers, enabling independent scaling of compute resources. StarRocks uses columnar storage and incorporates a pipelined execution engine. Version 3.1.4 is used in this research.

The systems are selected based on their architectural diversity and documented designs. DuckDB and MonetDB provide open-source implementations with published academic literature. While Hyper is closed-source, its architecture is documented in research papers. StarRocks offers open-source implementation with technical documentation. These systems share common analytical processing capabilities while implementing different architectural approaches, enabling comparative analysis of how various design decisions affect environmental impact. Table \ref{tab:databases} presents detailed technical specifications of each system.

\begin{table*}[h]
\centering
\caption{Comparison of Database Systems}
\begin{tabular}{|l|c|c|c|c|}
\hline
\textbf{Database} & \textbf{DuckDB} & \textbf{Hyper} & \textbf{MonetDB} & \textbf{StarRocks} \\
\hhline{|=|=|=|=|=|}
Scalability & Vertical & Vertical & Vertical & Vertical/Horizontal \\
\hline
Storage Architecture & Hybrid & Hybrid & Hybrid & Disk-oriented \\
\hline
Storage Model & Columnar & Hybrid & Columnar & Columnar \\
\hline
Query Compilation & Precompiled Primitives/Vectorized model & Just-In-Time & Precompiled Primitives & Just-In-Time \\
\hline
Concurrency & MVCC/Optimistic concurrency control  & MVCC & Optimistic concurrency control & MVCC \\
\hline
Use Case & OLAP & OLAP/OLTP & OLAP & OLAP \\
\hline
Open Source & Yes & No & Yes & Yes \\
\hline
License & MIT & Proprietary & Apache 2.0 & Apache 2.0 \\
\hline
Developer & CWI & Tableau/TUM & CWI & StarRocks Inc. \\
\hline
First Released & 2019 & 2010 & 2004 & 2020 \\
\hline
\end{tabular}
\label{tab:databases}
\end{table*}

 \subsection{Energy Profiler}\label{subsec:rapl}
 Intel RAPL (Running Average Power Limit)~\cite{intel_sdm} is a power management technology integrated into Intel processors and provides a mechanism for monitoring and controlling the power consumption of a processor and some associated components like memory and graphics processor. Intel RAPL provides a platform-independent way to measure the energy consumed by different processor components, integrating power measurement circuits directly into the processor. This allows the software to monitor power consumption in real-time. It provides energy consumption in micro Joules natively, meaning no additional logic is needed to work out the amount of energy consumed by a processor while it is running.

 \subsection{Hardware Setup}
The experiments were conducted on a server with an 8-core Intel(R) Xeon(R) CPU E5-2637 v4 @ 3.50 GHz, featuring 2 sockets with 4 cores per socket and 2 threads per core, resulting in a total of 16 threads for parallel processing. The server includes DDR4 memory with 16 sockets totaling 512 GB of RAM along with two storage disks, a 3.5-inch HDD (1 TB), and a 2.5-inch SSD (512 GB).
 
\section{Environmental Metrics}

To assess the environmental impact of the examined database systems, we analyze two key perspectives: carbon emissions and water footprint. For carbon emissions, we employ the Software Carbon Intensity (SCI) specification~\cite{green_software_sci}, recommended by the Green Software Foundation as a standard metric for the carbon footprint of software. The SCI metric considers both operational carbon emissions (O) and the embodied emissions of the hardware (M), and it calculates the carbon footprint per application-specific unit of work (R): SCI = (O + M) per R.

Beyond carbon emissions, water footprint represents a crucial yet often overlooked environmental metric for software operations. Based on the Water Footprint Assessment Manual~\cite{hoekstra2011water}, we measure the total freshwater volume (surface or groundwater, i.e., blue water~\cite{schneider2013three}) used in both database operations and hardware manufacturing. We characterise the first category as operational water footprint and the second one as manufacturing water footprint. The operational water footprint pertains to the water associated with the electricity used to power the hardware, while the manufacturing water footprint concerns the water consumed during the production of the hardware components.

In the rest of this section, we provide a detailed explanation of each environmental metric and present our methodology for utilizing these metrics to measure the environmental impact of the examined database systems.

\subsection{Operational Carbon Emissions}\label{subsec:operational_carbon}

Our study focuses on calculating the operational carbon emissions from database servers, specifically examining the energy consumption of CPU and DRAM, which are the primary energy consumers in database systems~\cite{kumar2011memory}. We exclude energy consumption from cooling systems, GPUs, and other peripherals from our analysis. GPUs are rarely used in mainstream database systems as they are primarily optimized for specialized workloads like machine learning rather than traditional database operations. Additionally, the energy consumption of peripheral components represents a small fraction of a database server's total power draw compared to the dominant CPU and memory subsystems. The operational carbon emissions are derived from the electricity used to power the hardware, calculated using the formula: $O = I_E \times E_{op}$, where $O$ is the operational carbon emissions, $I_E$ is the grid carbon intensity (\si{gCO2/kWh}), and $E_{op}$ is the energy consumed (kWh).

This formula multiplies the total energy consumed by the carbon intensity of the energy source, which represents the amount of carbon dioxide emitted per kilowatt-hour of electricity produced. Carbon intensity depends on the local energy mix, varying by region and power source (e.g., gas at 490 \si{gCO2/kWh} compared to solar at 41 \si{gCO2/kWh}~\cite{gupta2022act}), making a database server's geographical location crucial for its operational emissions. We source regional intensity data from Electricity Maps~\cite{electricitymap}.

\subsection{Embodied Carbon Emissions}

Embodied carbon emissions refer to the total greenhouse gases (GHGs) emitted throughout the manufacturing cycle of a product, excluding its operational use. In the context of IT infrastructure, these emissions encompass all the carbon released during the extraction and processing of raw materials, the manufacturing and assembly of hardware components, and the transportation of these components to their final destination.

In this paper, we focus on the embodied carbon emissions related to the manufacturing process of each individual component of a database server (CPU, DRAM, storage medium) and the packaging of these hardware components into a single machine. The total embodied carbon emissions of the hardware components can be obtained from published papers~\cite{gupta2022act} and other resources such as hardware vendors~\cite{skhynix2020,skhynix_carbon, seagatessd120, seagatehdd, seagate25}.

Table~\ref{tab:embodied_carbon} shows the embodied carbon footprint values that we are using.  Since direct data for our exact hardware components is unavailable, we leverage reported embodied carbon emissions from components with similar technologies.

\subsubsection{\textbf{CPU Carbon Emissions Calculation}}

For the CPU, we could not find data from the manufacturers, so we followed the methodology specified in~\cite{gupta2022act}: $M = \text{Area} \times \text{CPA}$, where Area is the CPU chip die area (246.24 mm² per socket in our case \cite{wikichip_xeon_e5}), and CPA (carbon per unit area) is calculated as follows:

\begin{equation}
    \text{CPA} = \frac{1}{Y} \times (\text{CI}_{\text{fab}} \times \text{EPA} + \text{GPA} + \text{MPA}),
\end{equation}

where $Y$ is the fab yield, $\text{EPA}$ is the energy consumed per unit area manufactured, $\text{GPA}$ is emissions per unit area from chemicals burned during hardware manufacturing, $\text{MPA}$ is the emissions from procuring raw materials for fab manufacturing, and $\text{CI}_{\text{fab}}$ is the carbon intensity of the fab that produced the CPU.

From Gupta et al.~\cite{gupta2022act}, we have:
$$\text{EPA} = 1.2~\si{kWh/cm^2}, \text{GPA} = 200~\si{gCO2/cm^2},  \text{MPA} = 500~\si{gCO2/cm^2}$$

For the fab yield, based on data published by Intel~\cite{moorhead2014intel} and data related to its competitors~\cite{wccftech2020smic,shilov2019smic} for 14nm lithography, we assume a fab yield of 90\%. Intel's 14nm CPU chips are manufactured at Leixlip, Ireland \cite{intel2023ireland}, using an energy grid of 78.3\% renewable, 21.6\% natural gas, and 0.1\% diesel \cite{intel2023climate}. Using wind as the renewable energy source and with carbon intensities from \cite{electricitymap} of 13 (wind), 594 (gas), and 885 (diesel) gCO$2$eq/kWh, we derive $\text{CI}_{\text{fab}}$ = 139.368 gCO$_2$eq/kWh. This yields CPA = 963.60 gCO$_2$/cm$^2$ and total emissions M = 2.46 × 963.60 = 2370.46 gCO$_2$ per socket.

\begin{table}[H]
\centering
\caption{Embodied Carbon Footprint}
\begin{tabular}{|l|c|}
\hline
\textbf{Type} & \textbf{Emissions} \\
\hhline{|=|=|}
DRAM DDR4 10nm & 63 gCO2e/GB~\cite{skhynix2020} \\
\hline
HDD 3.5 inch 1TB & 7.11gCO2e/GB~\cite{seagate25} \\
\hline
SSD TLC NAND Flash 512GB & 26.7gCO2e/GB~\cite{seagatessd120} \\
\hline
Intel(R) Xeon(R) CPU E5-2637  & 4.74 kgCO2e \\
\hline
\end{tabular}
\label{tab:embodied_carbon}
\end{table}

\subsection{Operational Water Footprint}\label{subsec:operational_water}
The operational water footprint in this study quantifies the blue water consumption associated with electricity generation for database server operations. Blue water refers to fresh surface and groundwater resources from rivers, lakes, and aquifers,encompassing both potable and non-potable freshwater. This consumption specifically measures the volume of water that was not returned to the original water source primarily due to evaporation, transpiration, and incorporation into the byproducts of the power generation. The calculation excludes green water (rainwater stored in soil moisture available for plant growth and evapotranspiration) and grey water (the freshwater volume needed to dilute pollutants to meet water quality standards), focusing solely on direct freshwater consumption in energy production for database operations. We focus specifically on blue water because its consumption is more directly measurable and consistently reported in the power generation literature, enabling more reliable analysis despite existing data gaps. Additionally, blue water consumption benefits from more established and standardized quantification methodologies, facilitating more reliable comparisons across different electricity generation technologies and regional contexts. For determining the database server's operational water consumption, we rely on current research \cite{jin2019water} on the water footprint of electricity. Table \ref{tab:op_water_footprint} presents the respective consumptive water footprint per unit of power, as presented in \cite{jin2019water}, for convenience of the reader.

\begin{table}[h]
\centering
\caption{Water Footprint of Various Energy Sources}
\begin{tabular}{|l|c|}
\hline
\textbf{Energy Source} & \textbf{Water Footprint (\si{L}/\si{MWh}) - Median} \\
\hhline{|=|=|}
Biomass/Firewood & 1817 \\
\hline
Hydropower & 51480 \\
\hline
Nuclear & 2200 \\
\hline
Oil & 1746 \\
\hline
Coal and Lignite & 1817 \\
\hline
Geothermal & 1363 \\
\hline
Natural Gas & 700 \\
\hline
Solar & 45 \\
\hline
Wind & 1.85 \\
\hline
\end{tabular}
\label{tab:op_water_footprint}
\end{table}

\begin{figure*}[!b]
    \centering
    \begin{minipage}{0.48\textwidth}
        \centering
        \includegraphics[width=\textwidth]{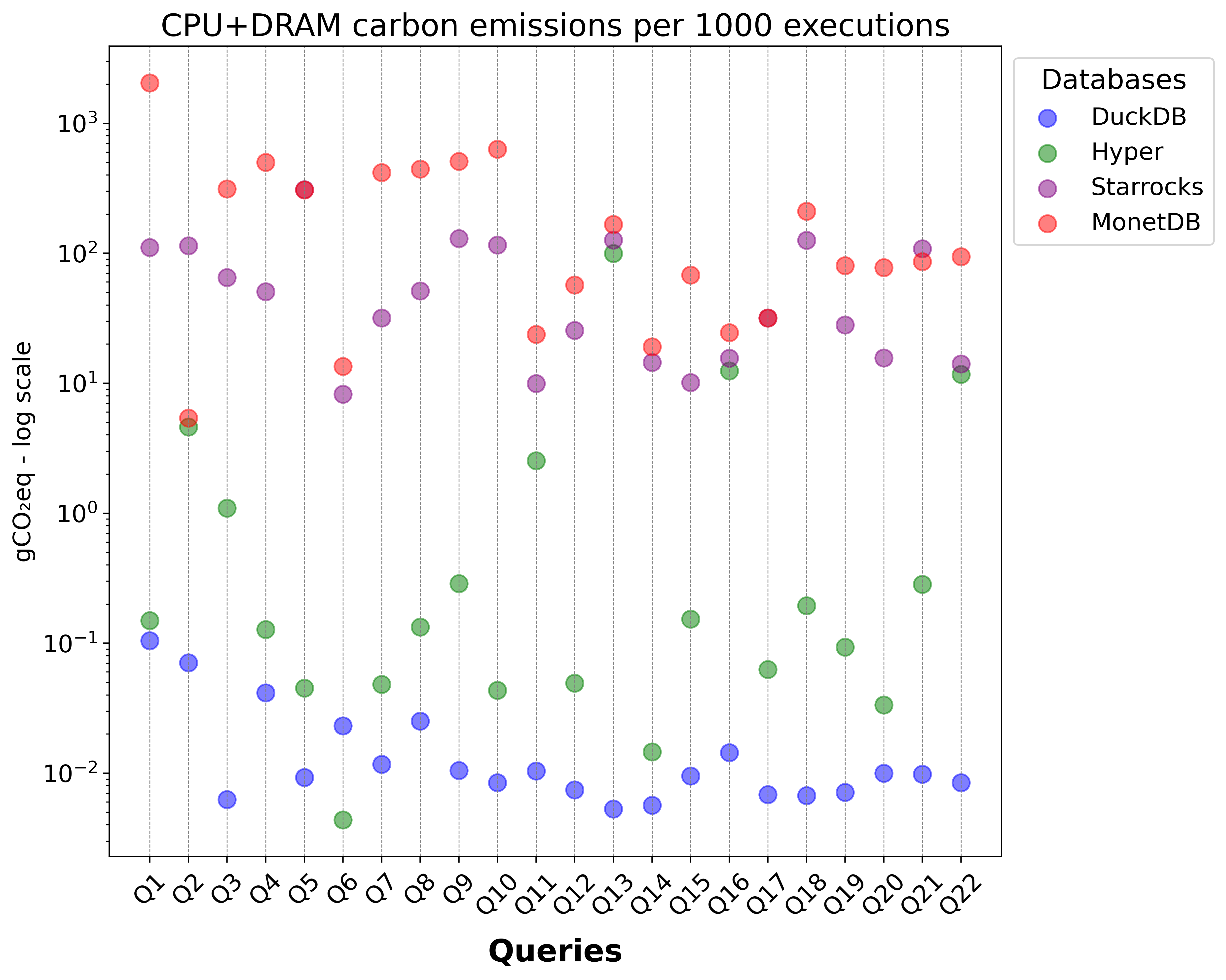}
        \caption*{(a) CPU and DRAM carbon emissions for TPC-H queries}
        \label{fig:cpu_carbon_tpch}
    \end{minipage}
    \quad
    \begin{minipage}{0.48\textwidth}
        \centering
        \includegraphics[width=\textwidth]{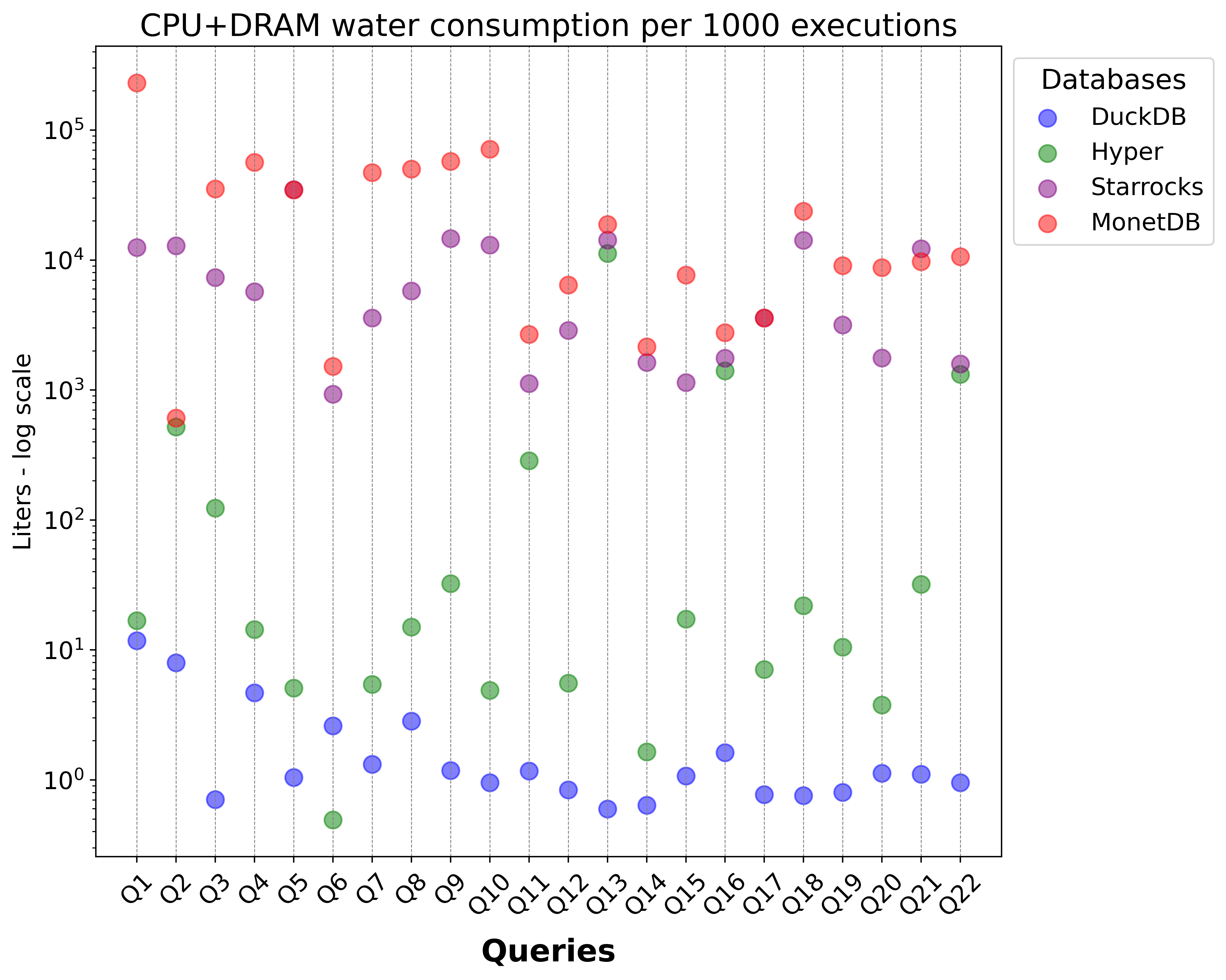}
        \caption*{(b) CPU and DRAM water consumption for TPC-H queries}
        \label{fig:cpu_water_tpch}
    \end{minipage}
    \caption{Environmental impact metrics for TPC-H queries}
    \label{fig:tpch_queries}
\end{figure*}

\subsection{Manufacturing Water Footprint}
The manufacturing water footprint refers to the total volume of freshwater used during the product's manufacturing process, including both direct and indirect water consumption. This includes water used in the production of electricity to power the semiconductor fabs as well as water used in the extraction and processing of raw materials, numerous manufacturing and assembly processes, and other activities required to produce the final product. It also accounts for the dirty water produced by these processes, which must be treated before being returned to the environment. For example, in electronics manufacturing, this footprint comprises water used for metal mining, silicon wafer production, component assembly, and wastewater treatment.

Table 4 presents the manufacturing water footprint values derived from sustainability reports that we use in our analysis for different hardware components. However, due to the lack of specific data for the manufacturing water footprint of CPUs, we adopt an alternative approach for this component, taking into account the die size and the values from Frost et al.~\cite{frost2019quantifying} for the volume of water needed for the production per cm² of wafer. In our case, each CPU socket has a die area of 2.4624 \(cm^2\). Thus, in order to produce each CPU socket, the manufacturing water footprint is 7.8 \(\frac{L}{cm^2}\) * 2.4624 \(cm^2\) = 19.188 L.

\begin{table}[h]
\centering
\caption{Manufacturing Water Footprint}
\begin{tabular}{|l|c|}
\hline
\textbf{Type} & \textbf{Water} \\
\hhline{|=|=|}
DRAM DDR4 10nm & 1.12 L/GB~\cite{skhynix2020} \\
\hline
HDD 3.5 inch 1TB & 0.17 L/GB~\cite{seagate25}  \\
\hline
SSD TLC NAND Flash 512GB & 0.096 L/GB~\cite{seagatessd120} \\
\hline
Intel(R) Xeon(R) CPU E5-2637  & 38.376 L \\
\hline
\end{tabular}
\label{tab:embodied_water}
\end{table}

\section{Environmental Assessment Framework}\label{sec:framework}
This paper introduces a comprehensive approach to evaluating the environmental impact of analytical database systems through two complementary components: EcoQuery, a measurement and benchmarking infrastructure, and ATLAS, a systematic methodology for environmental impact assessment. Together, these frameworks enable rigorous evaluation of database systems' environmental efficiency across different architectures, configurations, and deployment scenarios.

Our approach builds directly upon the environmental metrics established in Section 4, where we defined the theoretical foundation for quantifying database systems' environmental impact. The four key metrics—operational carbon emissions, embodied carbon emissions, operational water footprint, and manufacturing water footprint—are systematically implemented within our evaluation frameworks. EcoQuery provides the practical infrastructure to measure these metrics during database operations, while ATLAS offers the methodology to interpret and analyze them in meaningful contexts. Through this integration, we transform abstract environmental measures into actionable insights about database system sustainability.

\subsection{EcoQuery: Environmental Impact Measurement Infrastructure}\label{subsec:benchmark}
We introduce EcoQuery,  a comprehensive framework for measuring and evaluating the environmental impact of analytical database systems. Unlike traditional database benchmarks that focus solely on performance metrics, EcoQuery implements standardized methodologies for capturing both operational and manufacturing environmental impacts while maintaining compatibility with industry-standard workloads. To facilitate reproducibility and further research in this area, we have made the complete EcoQuery framework publicly available on GitHub (\url{https://github.com/MSRG/EcoQuery}).

The framework implements a measurement architecture designed to provide comprehensive environmental impact assessment while minimizing overhead. At its core, EcoQuery interfaces directly with Intel RAPL to collect precise CPU and DRAM energy measurements, while simultaneously gathering detailed I/O statistics and resource utilization data. This data collection process operates continuously during query execution, with temporal aggregation ensuring accurate correlation between database operations and their environmental impact.

EcoQuery extends traditional benchmarks like TPC-H through environmental instrumentation while maintaining their original performance metrics. For our research, we utilize TPC-H at 300GB scale factor as our reference workload. The benchmark's standardized schema of 8 tables representing a wholesale supplier business model and its 22 complex queries provide a comprehensive test of analytical processing capabilities while enabling consistent environmental impact measurements. 

The framework standardizes collection and reporting of both operational and manufacturing metrics, including energy consumption measurements for CPU, DRAM, along with derived metrics such as carbon emissions and water consumption. Manufacturing impact metrics encompass embodied carbon from hardware components, manufacturing water footprint, and considerations for hardware lifecycle and component replacement impacts. 

\subsection{ATLAS: A Systematic Methodology} \label{subsec:methodology}

Building upon EcoQuery's measurement capabilities, we present ATLAS (AnaLytical daTAbase Sustainability), a systematic methodology for evaluating the environmental impact of analytical database systems. ATLAS integrates operational and manufacturing environmental metrics to provide a complete assessment framework that considers carbon emissions from both operational energy consumption and hardware manufacturing, water consumption across power generation and hardware production processes, geographical variations in environmental impact, system lifecycle environmental efficiency, hardware configuration impacts, and storage medium environmental implications.

\subsubsection{\textbf{Operational Environmental Impact Analysis.}}\label{subsubsec:operational}\hspace{-9pt}
ATLAS focuses primarily on measuring operational environmental impacts using several critical metrics:

\paragraph{Energy Consumption}
The operational analysis begins with measuring CPU and DRAM energy consumption for each database system under controlled conditions using EcoQuery. These measurements serve as the foundational data for deriving additional environmental impact metrics, such as carbon emissions and water footprint, enabling a detailed understanding of energy usage across diverse database architectures and query processing mechanisms.

\paragraph{Derived Environmental Metrics: Carbon Emissions and Water Footprint}
From the measured energy consumption, operational carbon emissions are estimated by incorporating the carbon intensity of the power grid at the server's location. Additionally, the operational water footprint is calculated by combining the observed energy consumption with the water footprint of the electricity grid's energy sources~\cite{jin2019water} during the measurement periods.

\paragraph{Hardware Configuration Impact}
To further the analysis and evaluate environmental impacts under varying conditions, ATLAS includes experiments with different hardware configurations, such as varying memory sizes. This analysis quantifies the influence of resource allocation on carbon emissions and water consumption, offering insights into the environmental trade-offs of database system deployment and resource optimization strategies from the environmental perspective.

\paragraph{Geographical and Regional Energy Impact Analysis}
Building on the derived environmental metrics, ATLAS incorporates a geographical perspective to examine how variations in regional energy mixes influence operational carbon emissions. By considering the carbon intensity of energy sources specific to each region, this analysis underscores the critical role of server location in shaping environmental impacts. These insights enable more informed and environmentally conscious deployment decisions, particularly for distributed database systems.

\paragraph{Storage Medium Analysis}
Separately, ATLAS evaluates the environmental implications of storage medium choices, such as HDD and SSD, on CPU-related energy consumption and carbon emissions. This analysis highlights the interplay between storage technology and overall system efficiency, providing guidance on optimizing storage configurations for reduced environmental impact.

\begin{figure*}[b!]
    \centering
    \begin{minipage}{0.47\textwidth}
        \centering
        \includegraphics[scale=1.5, width=\textwidth]{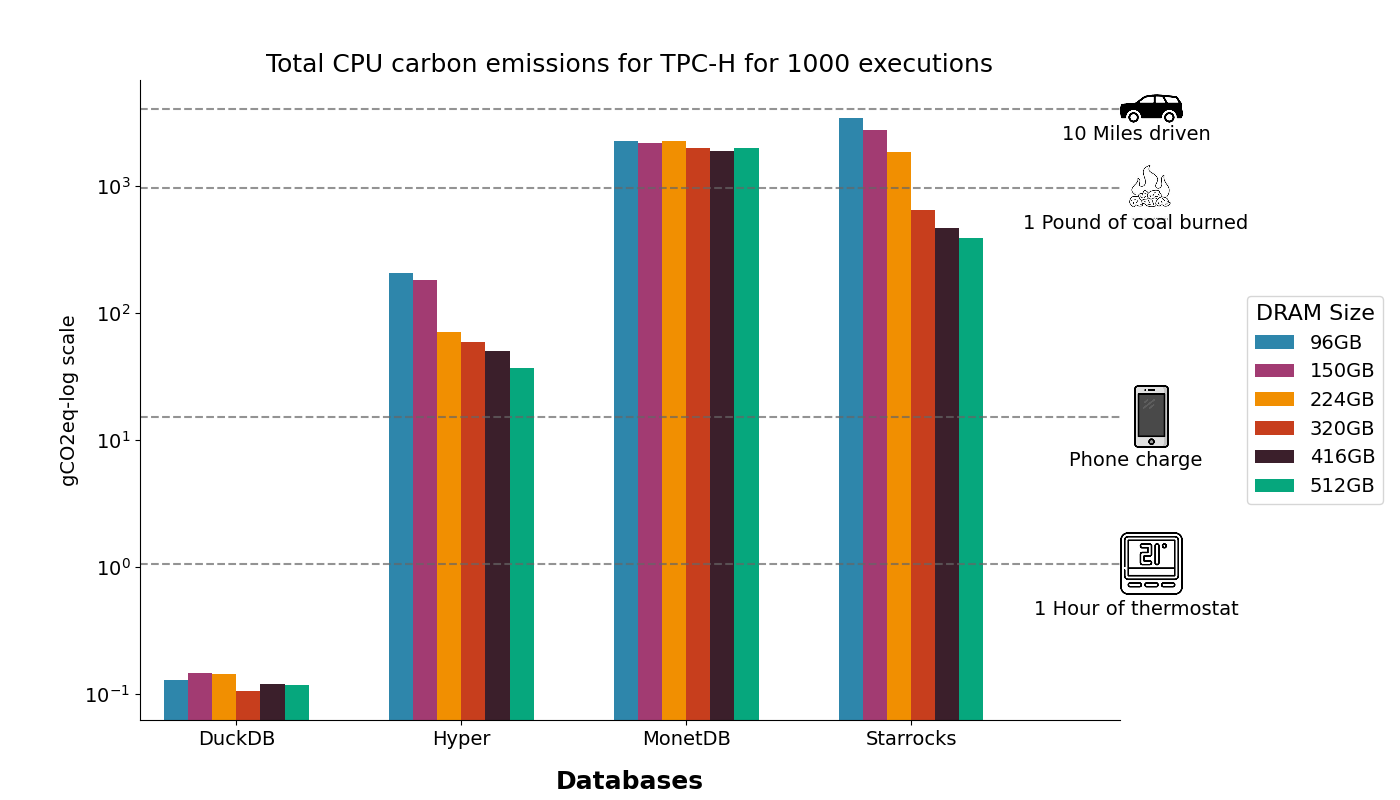}
        \caption*{(a) CPU carbon emissions for TPC-H queries}
        \label{fig:cpu_carbon_dram}
    \end{minipage}
    \quad
    \begin{minipage}{0.47\textwidth}
        \centering
        \includegraphics[scale=1.5, width=\textwidth]{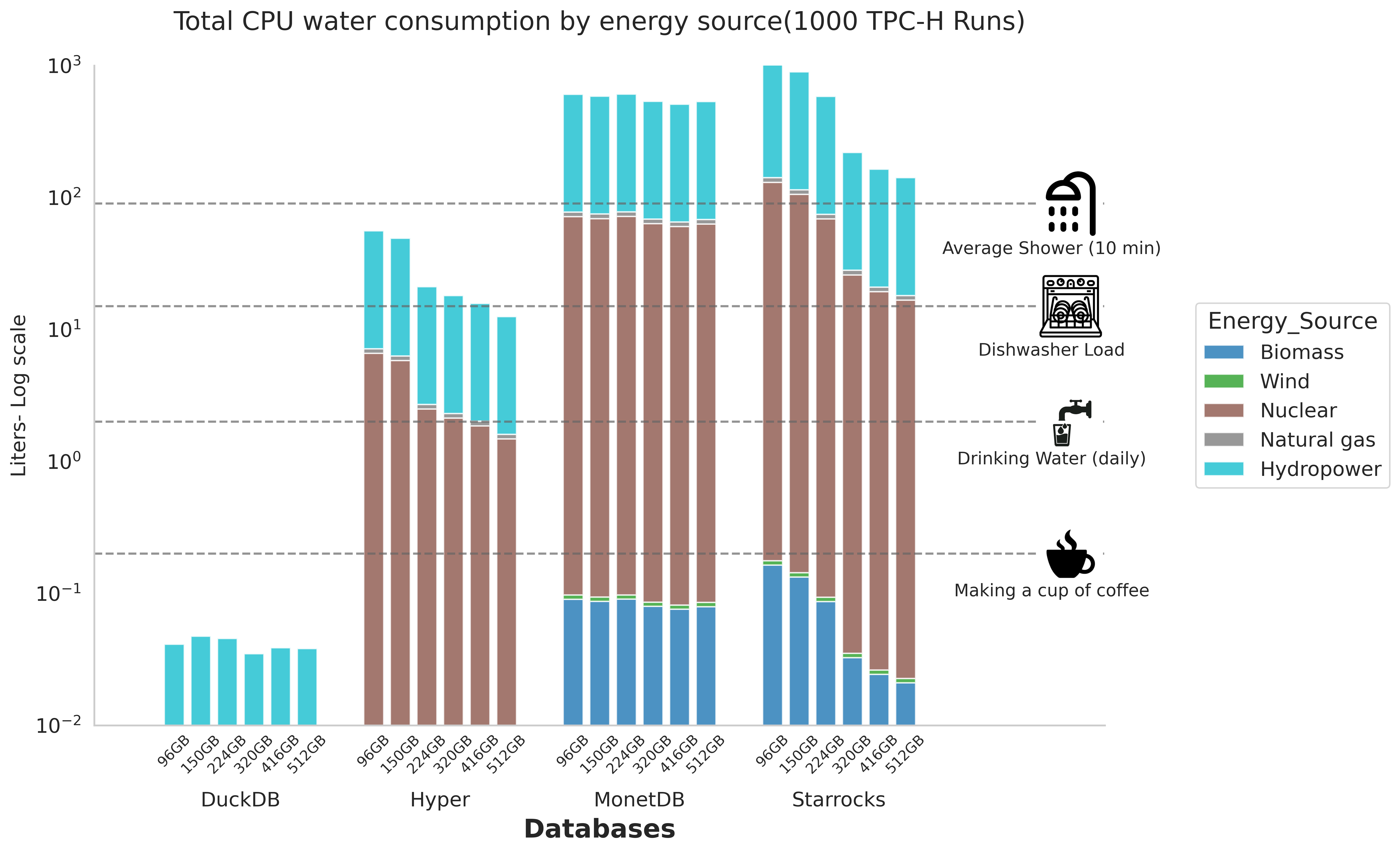}
        \caption*{(b) CPU water consumption for TPC-H queries}
        \label{fig:cpu_water_dram}
    \end{minipage}
    \caption{Environmental impact metrics for different DRAM sizes}
    \label{fig:changing_dram}
\end{figure*}

\subsubsection{\textbf{Manufacturing Environmental Impact Analysis}}\label{subsubsec:manufacturing}\hspace{-9pt}
In addition to operational impacts, ATLAS extends to manufacturing environmental considerations:

\paragraph{Break-even Analysis}
ATLAS conducts a break-even analysis to evaluate the balance between operational and embodied carbon emissions. This analysis estimates the number of queries and the time required to break-even manufacturing emissions, assessing whether this manufacturing emissions amortization point occurs within the server's expected lifespan. Such insights are essential for understanding the total environmental footprint of database deployment decisions and identifying optimal system lifetimes.

\paragraph{Power Efficiency Analysis}
To support the break-even analysis, ATLAS examines workload power efficiency at a granular level. Power efficiency directly influences energy consumption, carbon emissions, and water usage during query execution, determining how quickly the database systems approach the break-even point. For example, stable and efficient power consumption patterns enhance carbon efficiency, reducing the operational emissions required to amortize manufacturing emissions. Conversely, inefficient or variable power usage can extend the time needed to reach the break-even point, impacting the overall environmental performance of the database system. This analysis provides deeper insights into the energy-use dynamics that underpin the environmental sustainability of database operations.

\paragraph{Storage Medium Endurance}
Building on the break-even analysis, ATLAS investigates how different database architectures affect the operational lifetime of the underlying hardware, with the goal of maximizing the environmental value of the initial carbon investment. After a database server reaches the break-even point and moves into the post-amortization phase, the next step is to analyze how varying database architectures influence hardware durability over time. This aspect of the analysis evaluates storage medium endurance through an analysis of I/O operations during workload execution, investigating how query processing patterns influence storage wear and assessing the long-term environmental implications of architectural decisions. 
Furthermore, it links storage durability to embodied carbon emissions and manufacturing water footprints, highlighting the environmental significance of storage-related architectural choices. In this way, ATLAS provides insights into how different architectures can optimize hardware usage, leading to a reduction in both carbon and water footprints beyond the break-even point.

This multifaceted methodology offers a detailed framework for understanding the environmental implications of database system selection, configuration, and deployment. By addressing both immediate operational impacts and long-term manufacturing considerations, it provides a robust foundation for making environmentally informed decisions in database system design and deployment.

\begin{figure*}[b!]
    \centering
    \begin{minipage}{0.47\textwidth}
        \centering
        \includegraphics[scale=1.5, width=\textwidth]{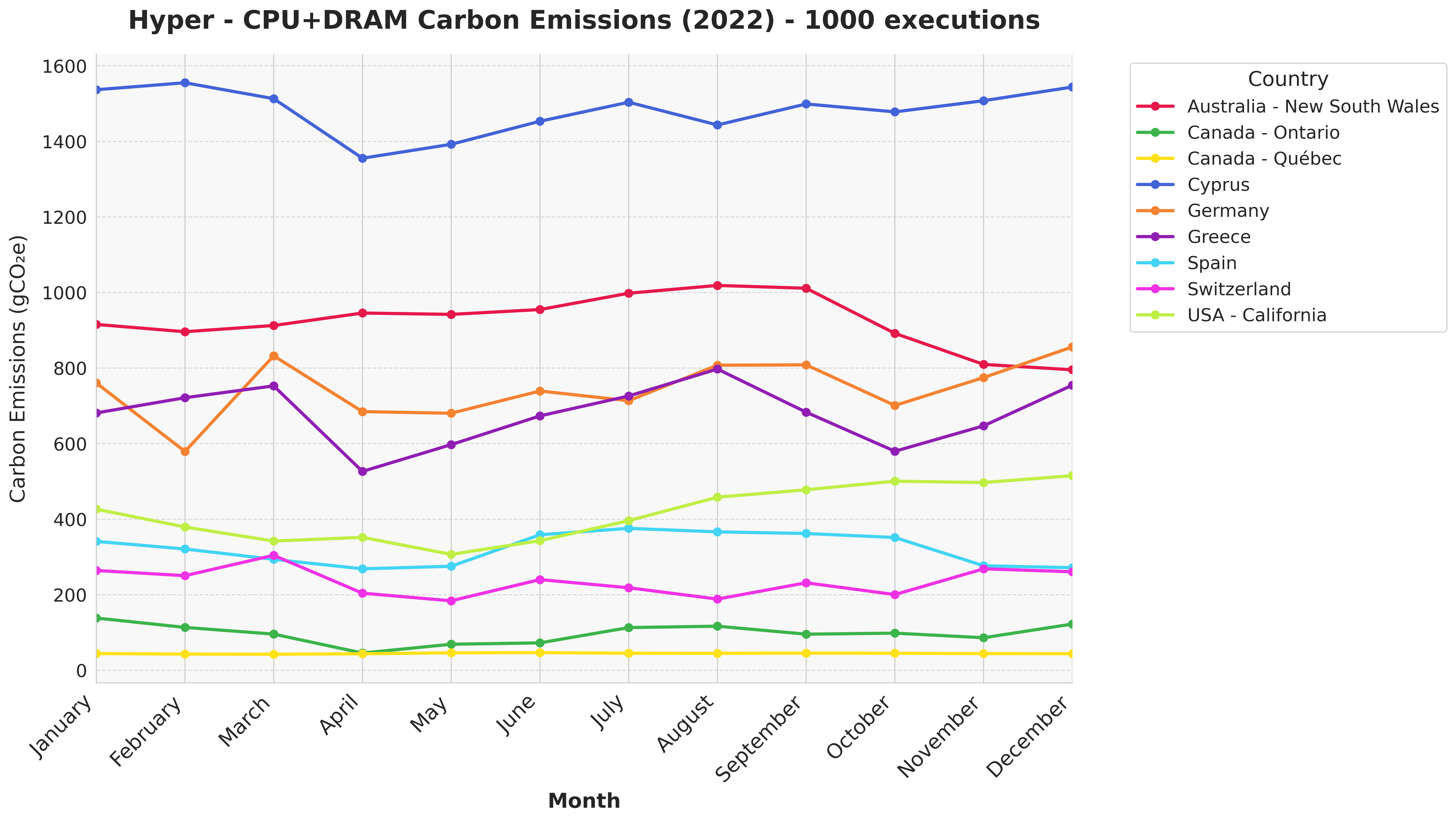}
        \caption*{(a) Hyper carbon emissions across 2022 in different locations}
        \label{fig:temporal_emissions_hyper}
    \end{minipage}
    \quad
    \begin{minipage}{0.47\textwidth}
        \centering
        \includegraphics[scale=1.5, width=\textwidth]{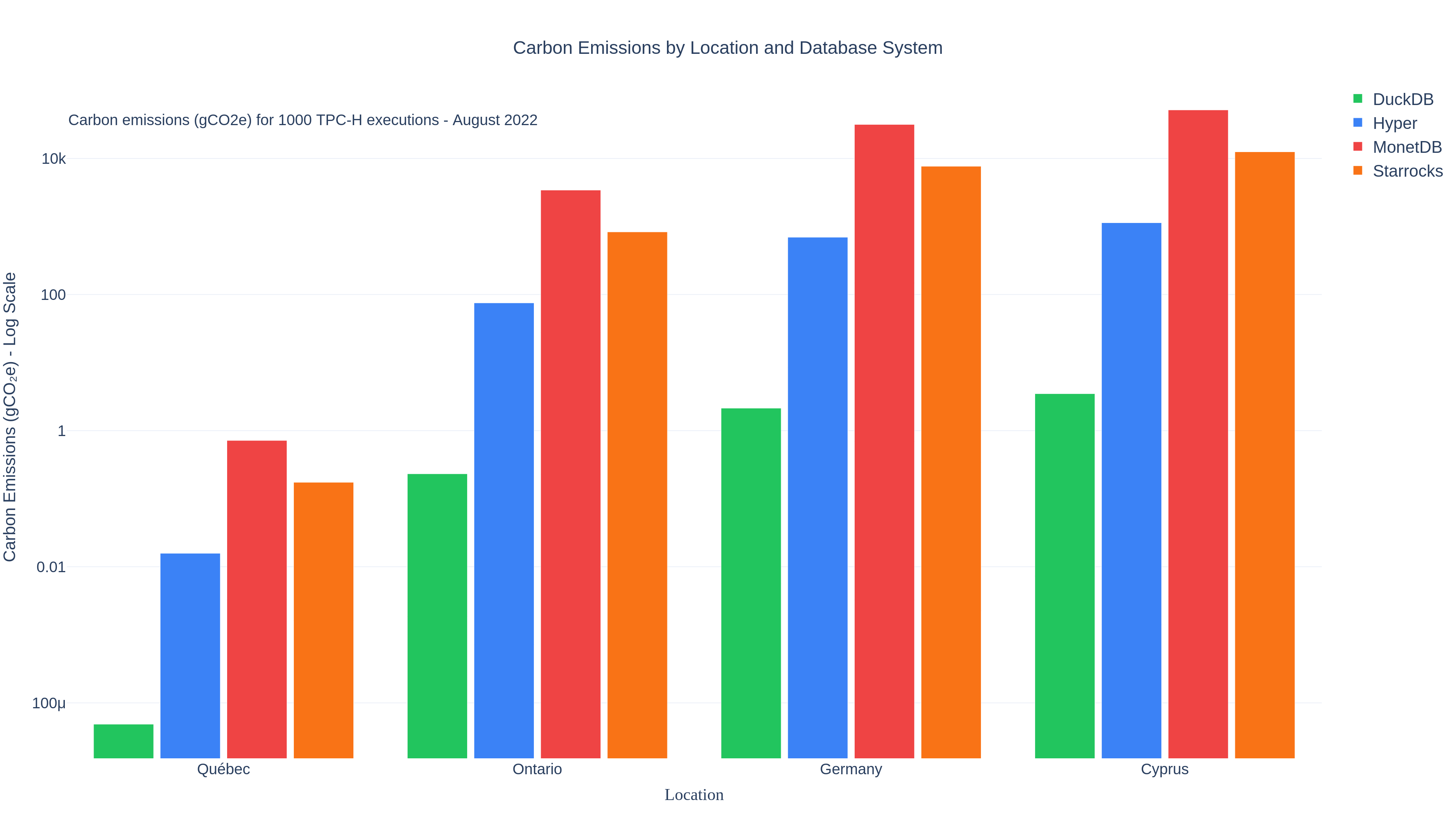}
        \caption*{(b) Carbon Emissions by Location and DBMS(August 2022)}
        \label{fig:location_emissions}
    \end{minipage}
    \caption{Temporal and Regional Carbon Emissions Analysis}
    \label{fig:temporal_regional_emissions}
\end{figure*}

\section{Application of ATLAS: Experimental Evaluation and Analysis}

In this section, we systematically apply the ATLAS methodology to evaluate the environmental impact of analytical database systems using our EcoQuery benchmark framework. Our experimental evaluation follows the structured methodology introduced in Section ~\ref{subsec:methodology}, examining both operational and manufacturing environmental impacts. The methodology guides our investigation through two main phases: operational environmental impact analysis and manufacturing environmental impact analysis. Each phase implements specific components of the ATLAS framework to ensure a comprehensive evaluation of database environmental efficiency.

\subsection{Operational Environmental Impact Analysis}

Following ATLAS's operational framework (Section~\ref{subsubsec:operational}), we analyze query execution's environmental impact across database systems. For energy consumption measurement, we employ Intel RAPL (Section~\ref{subsec:rapl}) to collect precise CPU and DRAM energy measurements during query execution. These measurements are then processed to calculate the environmental impact, by applying Equation (1) from Section~\ref{subsec:operational_carbon} for carbon emissions and utilizing the water footprint methodology from Section~\ref{subsec:operational_water}, incorporating regional grid variations from Electricity Maps~\cite{electricitymap}. This methodological implementation allows us to systematically evaluate both carbon emissions and water consumption across different query types, hardware configurations, and geographical locations.

\subsubsection{\textbf{Per-Query Environmental Impact Analysis}}\hspace{-9pt}
We begin the operational analysis by quantifying how different database architectures impact environmental efficiency through comprehensive measurements of EcoQuery's standardized TPC-H workload. To ensure a fair comparison, we measure both carbon emissions and water consumption per 1,000 query executions, providing a standardized basis for evaluation across diverse architectural approaches.

Figure~\ref{fig:tpch_queries}a illustrates the CPU and DRAM carbon emissions for all examined database systems across all TPC-H queries. DuckDB's embedded architecture and optimized in-memory processing yield the lowest emissions (0.01-0.1 gCO\textsubscript{2} per 1,000 executions), while MonetDB's operator-at-a-time model and extensive intermediate results generation produces the highest (exceeding 1,000 gCO\textsubscript{2} per 1,000 executions in Q1). Between these extremes, Hyper achieves intermediate efficiency, with StarRocks showing moderately higher emissions while remaining more efficient than MonetDB.

Water consumption patterns, illustrated in Figure~\ref{fig:tpch_queries}b, strongly correlate with carbon emission trends. DuckDB maintains its efficiency advantage, consuming around 1 liter of water per 1,000 executions across most queries. MonetDB's water consumption reaches up to $10^{5}$ liters per 1,000 executions for complex queries such as Q1, reflecting its energy-intensive execution model. Hyper and StarRocks maintain their intermediate positions, with their relative efficiency remaining consistent across both metrics.

The strong correlation between carbon emissions and water consumption patterns across database systems reflects a fundamental relationship in our current energy infrastructure: both metrics are proportional to energy consumption, which in turn correlates with query execution time. The striking similarity between the patterns in Figures~\ref{fig:tpch_queries}a and ~\ref{fig:tpch_queries}b is a direct consequence of our methodology. As detailed in Sections ~\ref{subsec:operational_carbon} and ~\ref{subsec:operational_water}, both carbon emissions and water consumption metrics are derived from the same underlying energy consumption measurements, with each metric applying different conversion factors based on the local electricity grid composition. This proportionality not only explains why the relative differences between database systems remain consistent across both figures, but also means that architectural decisions that improve energy efficiency naturally translate to environmental benefits across both metrics. Our analysis demonstrates this clearly as DuckDB's embedded architecture consistently outperforms client-server architectures like StarRocks and MonetDB across both metrics, most notably in computation-intensive queries like Q1, where it achieves emissions two orders of magnitude lower than MonetDB's approach. Hyper's hybrid architecture demonstrates that balanced design choices can achieve moderate environmental impact while maintaining system flexibility. The consistency of these patterns across both carbon emissions and water consumption suggests that environmental efficiency is inherently tied to fundamental system design decisions.

\subsubsection{\textbf{Hardware Configuration Impact Analysis}}\hspace{-9pt}
Following ATLAS's hardware configuration framework, we focus on memory, a critical aspect of database design that influences both performance and environmental efficiency. Available memory affects how database systems process queries, potentially leading to different execution strategies and varying levels of CPU utilization. Moreover, the interplay between memory size and architectural design choices can substantially impact environmental metrics. To investigate these relationships systematically, we measure the energy consumption, varying the server's DRAM size from 96GB to 512GB, while executing the complete TPC-H workload, and calculate the corresponding carbon emissions and water consumption.

\begin{figure*}[b!]
    \centering
    \begin{minipage}{0.48\textwidth} 
        \centering
        \includegraphics[width=\linewidth]{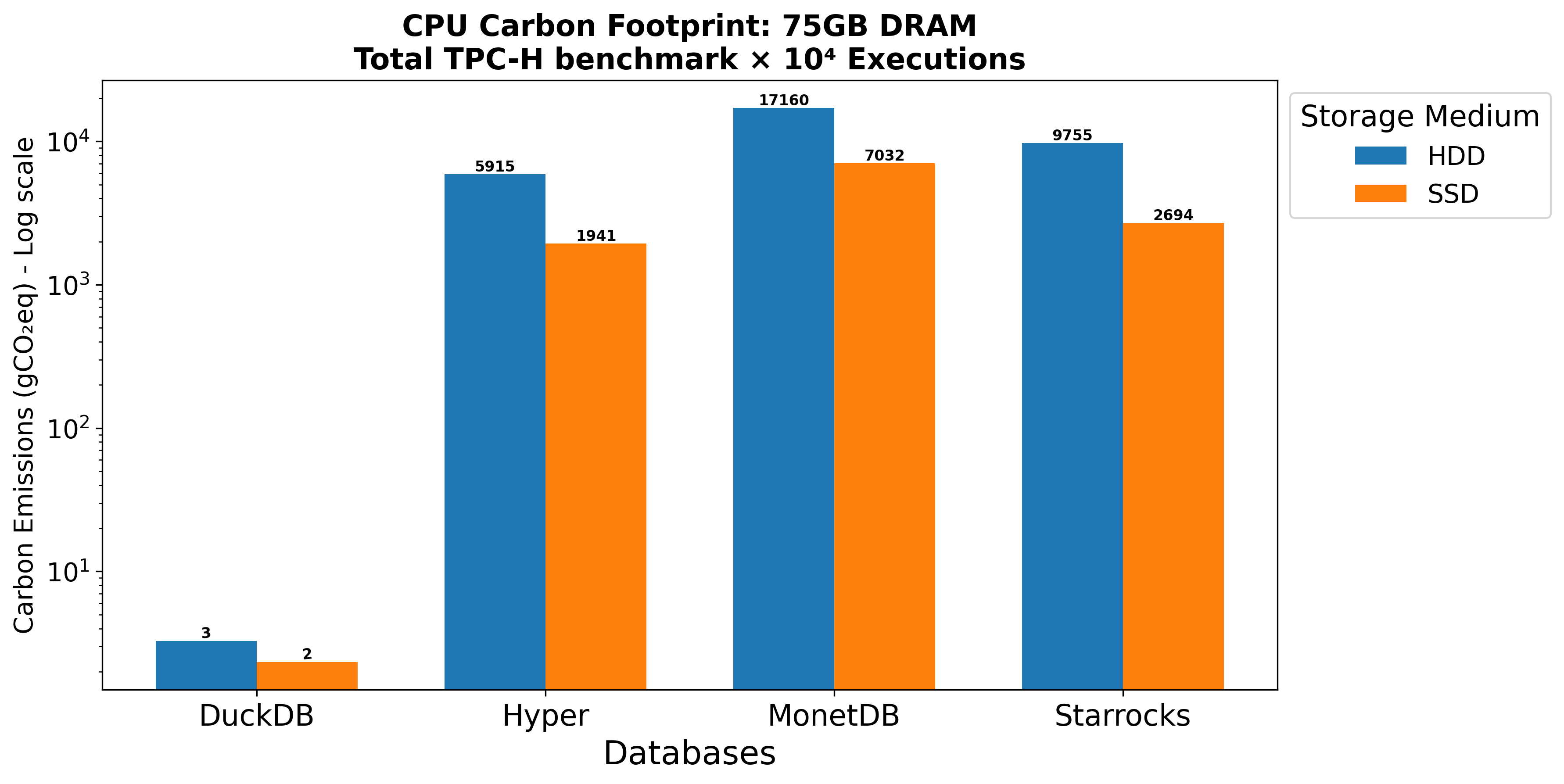}
        \caption*{(a) CPU Carbon Emissions by storage medium and DBMS}
        \label{fig:cpu_carbon_tpch_storage}
    \end{minipage}
    \hspace{0.02\textwidth} 
    \begin{minipage}{0.48\textwidth} 
        \centering
        \includegraphics[width=\linewidth]{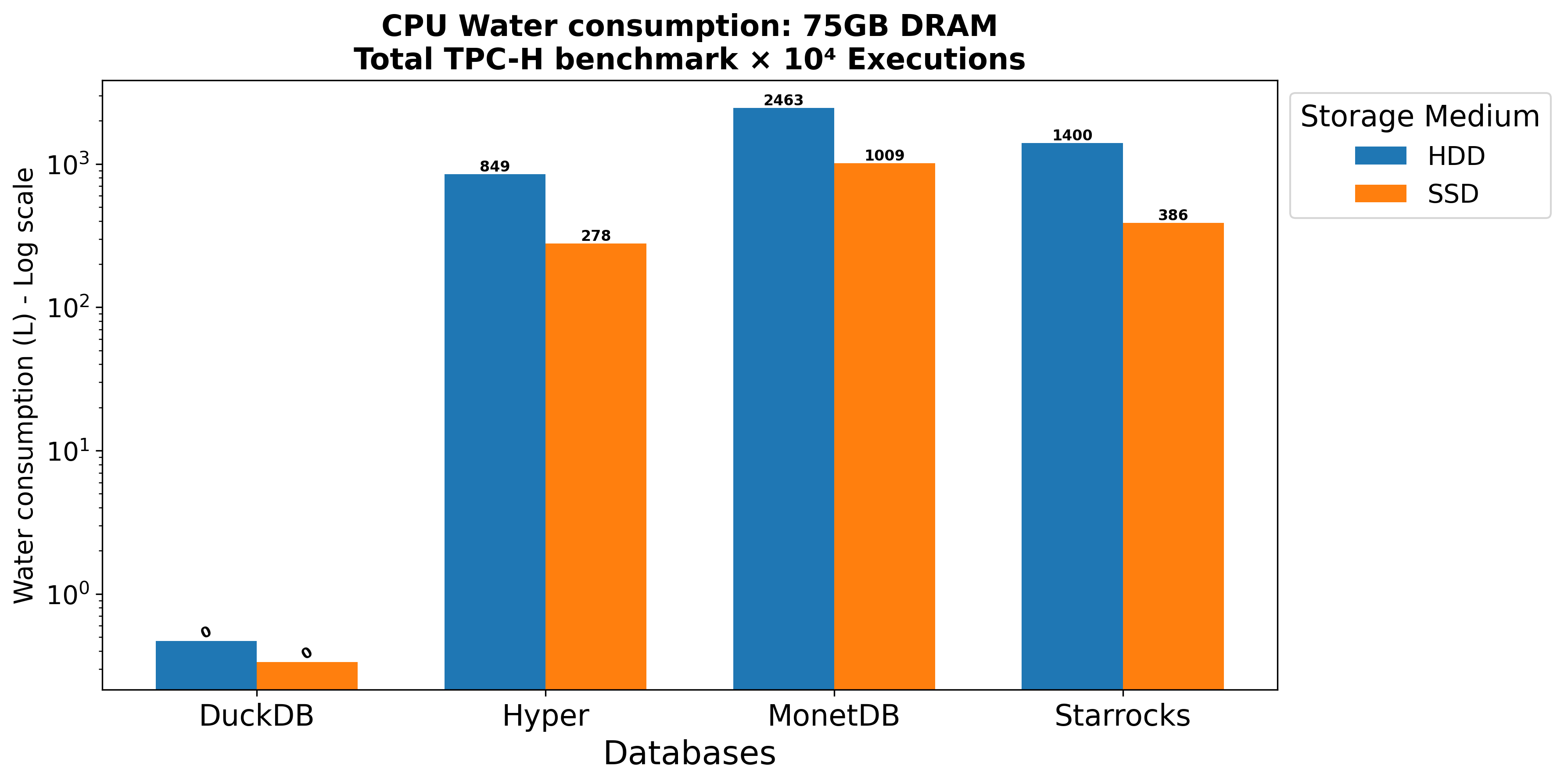}
        \caption*{(b) CPU Water footprint by storage medium and DBMS}
        \label{fig:cpu_water_tpch_storage}
    \end{minipage}
    \caption{Environmental impact metrics for different storage mediums}
    \label{fig:storage}
\end{figure*}

Figure~\ref{fig:changing_dram}a presents the total CPU carbon emissions for 1,000 executions TPC-H executions across DRAM configurations. For context, we include reference points from everyday activities. DuckDB's total emissions (approximately 0.1-0.2 gCO\textsubscript{2}) remain well below the carbon footprint of a single phone charge, while MonetDB and StarRocks' emissions (around 1,000-2,000 gCO\textsubscript{2}) approach the carbon impact of driving 10 miles or burning a pound of coal.

The results reveal that the memory's impact on environmental efficiency varies across architectures. DuckDB maintains consistently low emissions across all memory configurations, demonstrating efficient memory utilization without significant environmental penalties. In contrast, MonetDB shows the highest emissions but achieves a notable reduction (approximately 25\%) with increased memory, suggesting that larger memory allocations can help mitigate its operator-at-a-time execution model's environmental impact.

Figure~\ref{fig:changing_dram}b illustrates the water consumption breakdown based on Ontario's energy grid composition during the experiments. The analysis reveals how different energy sources—biomass, wind, nuclear, natural gas, and hydropower—contribute to the total water footprint. DuckDB's water consumption remains below that of making a coffee across all configurations, while MonetDB and StarRocks approach the water consumption of a dishwasher load. Notably, hydropower, despite being a renewable energy source, contributes significantly to the water footprint due to reservoir evaporation and operational requirements.  In contrast, biomass barely registers in the water footprint for more efficient systems like DuckDB and Hyper, with its contribution becoming virtually imperceptible due to their overall low water consumption. This breakdown highlights how regional energy infrastructure directly influences the environmental impact of database operations.

These findings demonstrate that while memory allocation influences environmental efficiency, its impact varies substantially based on the database architecture. The significant variations between different architectural approaches, even under identical memory configurations, emphasize the importance of considering both hardware resources and system design when optimizing for environmental impact.

\begin{figure*}[b]
    \centering
    \begin{minipage}{0.30\textwidth} 
        \centering
        \includegraphics[width=\linewidth]{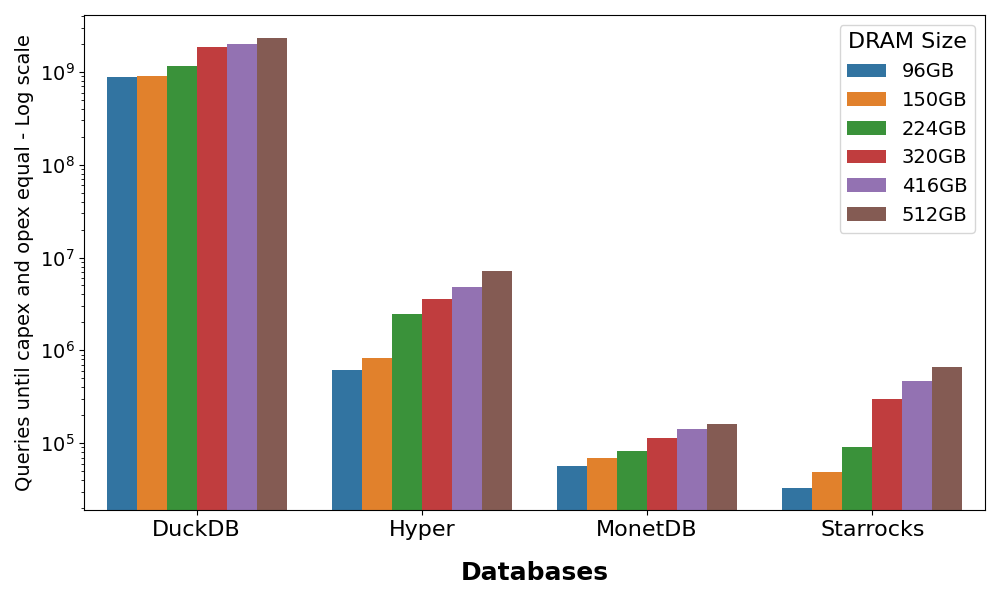}
        \caption*{(a) Number of queries until operational and embodied carbon emissions are equal}
        \label{fig:queries_breakeven}
    \end{minipage}
    \hspace{0.02\textwidth} 
    \begin{minipage}{0.30\textwidth}
        \centering
        \includegraphics[width=\linewidth]{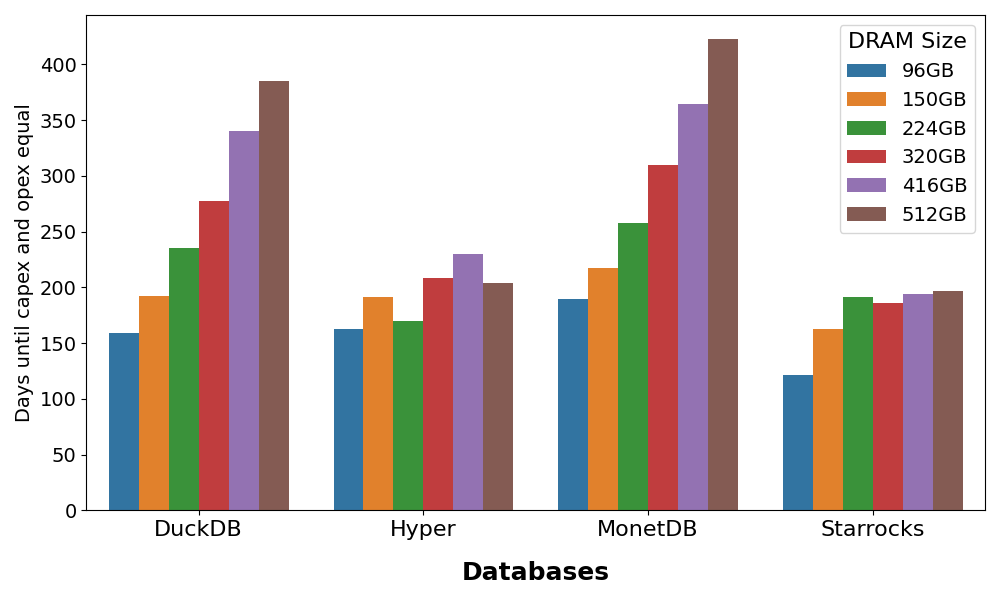}
        \caption*{(b) Days of continuous operation until operational and embodied carbon emissions are equal}
        \label{fig:days_breakeven}
    \end{minipage}
    \hspace{0.02\textwidth} 
    \begin{minipage}{0.31\textwidth}
        \centering
        \includegraphics[width=\linewidth]{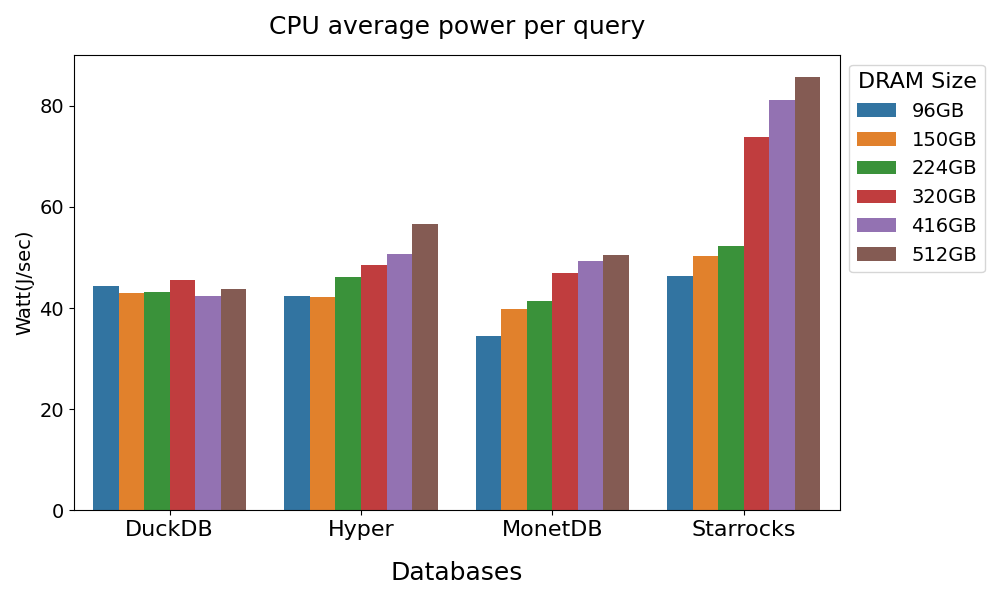}
        \vspace{1mm}
        \caption*{(c) Average power per query}
        \label{fig:power_efficiency}
    \end{minipage}
    \caption{Carbon efficiency - Break-even points and average power per query}
    \label{fig:break-even-comparison}
\end{figure*}

\subsubsection{\textbf{Geographical and Temporal Analysis}}\hspace{-9pt}
Our prior analyses focus on the operational environmental impact of various hardware configurations using Ontario's energy grid at a specific point in time. However, ATLAS acknowledges that the environmental impact of database operations can vary considerably depending on temporal and geographical factors. These variations arise from differences in regional energy sources and seasonal shifts in energy production patterns. To explore these effects, we analyze energy grids across multiple locations and timeframes with a consistent hardware configuration of 512GB memory~\cite{electricitymap}.

Figure~\ref{fig:temporal_regional_emissions}a presents a temporal analysis of Hyper's operational carbon emissions throughout 2022 across various geographical locations. The data reveals distinct patterns in different regions. Locations with high renewable energy penetration, such as Québec and Ontario, demonstrate relatively stable emission patterns throughout the year, with Québec consistently showing the lowest emissions due to its hydroelectric-dominated grid. In contrast, regions reliant on fossil fuels, such as Cyprus and New South Wales in Australia, dsiplay higher baseline emissions and greater temporal variability, attributed to seasonal changes in energy demand, availability of different power sources, and grid management strategies.

The impact of geographical location becomes even more pronounced when considering database systems with different energy efficiency profiles, as shown in Figure~\ref{fig:temporal_regional_emissions}b. Our analysis across Québec, Ontario, Germany, and Cyprus reveals that more energy-intensive architectures magnify the effect of location-based variations. For instance, while DuckDB's efficient architecture results in small absolute differences across locations (ranging from \SI{50}{\micro\gram\of{\ce{CO2}}} in Québec to 2gCO\textsubscript{2} in Cyprus), MonetDB's higher energy requirements lead to much larger differences (from 0.8gCO\textsubscript{2} to 20,000gCO\textsubscript{2} across the same locations), a five-orders magnitude difference for identical operations. This demonstrates how architectural efficiency interacts with geographical location to determine environmental impact.

These findings emphasize that the environmental impact of database operations must be considered within the context of both temporal and geographical factors. While efficient database architectures, as observed in our previous analyses, can significantly reduce their environmental impact, deployment location can amplify or diminish these architectural advantages.

\subsubsection{\textbf{Storage Medium Impact}}\hspace{-9pt}
Our analyses thus far have examined architectural differences, memory configurations, and geographical factors using HDD storage. As identified in ATLAS, storage medium selection represents another crucial decision that can influence environmental impact. To quantify these effects, we measure CPU energy consumption for both HDD and SSD storage across all database systems, and calculate the corresponding carbon emissions and water footprints based on the formulas in Sections ~\ref{subsec:operational_carbon} and \ref{subsec:operational_water}. We use a scale factor of 100GB, performed $10^{4}$ executions, and configured each system with 75GB of DRAM.

Figure~\ref{fig:storage}a illustrates the CPU carbon emissions for each database system under both storage configurations. SSD demonstrates operational environmental benefits across all architectures, with varying improvement magnitudes. DuckDB shows the smallest reduction (3.27 to 2.33 gCO2, HDD to SSD), reflecting its efficient embedded architecture and I/O optimization that perform well regardless of storage medium. In contrast, MonetDB exhibits the largest improvement, reducing from 17 kgCO2 to 7 kgCO2 (59\%). Hyper and StarRocks show intermediate improvements (67\% and 72\%, respectively), indicating effective utilization of SSD's I/O capabilities.

The water consumption patterns, shown in Figure ~\ref{fig:storage}b, closely mirror the carbon emission trends. DuckDB's water consumption decreases from 0.47 liters with HDD to 0.33 liters with SSD, demonstrating again the minimal environmental impact of its architecture across storage configurations. The most significant change appears in MonetDB's operations, where water consumption drops dramatically from 2,463 liters to 1,009 liters when switching to SSD. This dramatic improvement aligns with our understanding of MonetDB's operator-at-a-time execution model - its storage-intensive operations benefit substantially from SSD's reduced access latency and higher throughput. The consistent pattern of improvement across both metrics reflects their fundamental relationship with energy consumption, as both carbon emissions and water consumption scale proportionally with the energy required for database operations. This relationship indicates that storage medium selection significantly influences overall environmental efficiency, particularly for architectures with intensive I/O patterns.

These findings highlight how storage medium selection impacts database operations' environmental efficiency, with the magnitude of improvement varying based on architectural design. While SSDs consistently offer environmental benefits across all tested systems, the results suggest that more storage-intensive architectures stand to gain the most from solid-state storage adoption.

\begin{figure*}[b!]
    \centering
    \begin{minipage}{0.48\textwidth} 
        \centering
        \includegraphics[width=\linewidth]{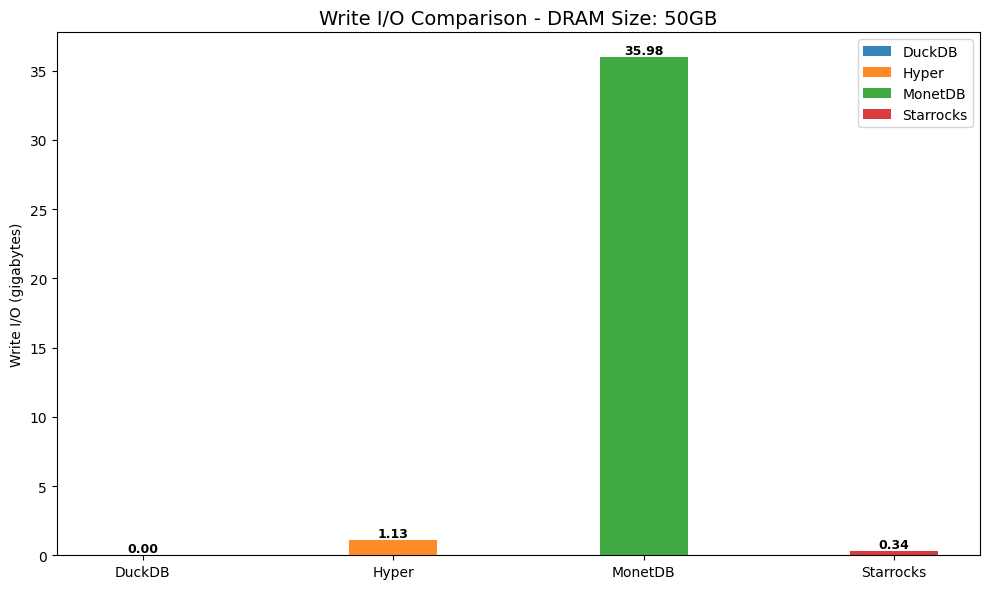}
        \caption*{(a) Write I/0 footprint for a single execution}
        \label{fig:wrtie_bytes_single_execution}
    \end{minipage}
    \hspace{0.02\textwidth} 
    \begin{minipage}{0.48\textwidth} 
        \centering
        \includegraphics[width=\linewidth]{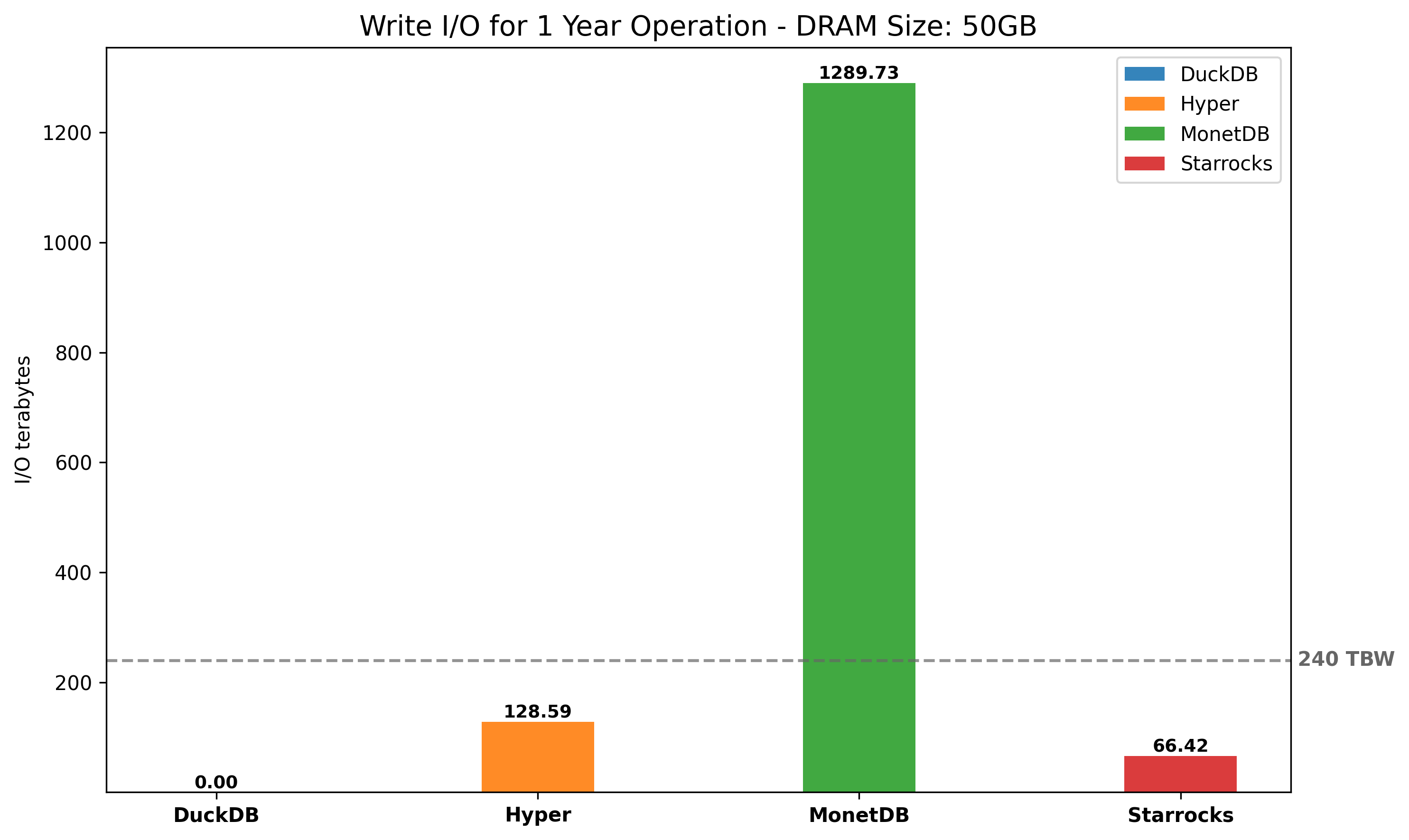}
        \caption*{(b) Write I/O foot for 1 year of continuous operation}
        \label{fig:wrtie_bytes_yearly_execution}
    \end{minipage}
    \caption{Write I/O analysis}
    \label{fig:writeio_footprint}
\end{figure*}

\subsection{Manufacturing Environmental Impact Analysis}

The long-term environmental implications of database deployments extend beyond operational impacts to include manufacturing costs and system lifetime efficiency. Our analysis implements the three core components of the ATLAS manufacturing methodology, as described in ~\ref{subsubsec:manufacturing}: break-even point determination, power efficiency evaluation, and hardware longevity assessment. Through this comprehensive approach, we uncover how architectural decisions influence both initial environmental investment and ongoing operational efficiency.

\subsubsection{\textbf{Break-even Analysis}}\hspace{-9pt}
Using ATLAS's break-even framework, we determine when the environmental cost of hardware manufacturing is amortized by the operational advantages.  The calculation incorporates two primary components: the total embodied carbon emissions from server components (CPU, memory, and storage) as specified in Table~\ref{tab:embodied_carbon}, and the operational emissions derived through the ATLAS operational framework detailed in Section~\ref{subsubsec:operational}. By examining the interplay between these factors, we can identify the point at which cumulative operational benefits justify the initial manufacturing investment.

Figure~\ref{fig:break-even-comparison}a illustrates the number of queries required to reach the carbon break-even point for each database system across different memory configurations. DuckDB demonstrates superior amortization of manufacturing emissions, requiring $10^{8}$ to $10^{9}$ queries to reach break-even. This high query requirement indicates that DuckDB's efficient operational profile  enables significantly more query processing before reaching manufacturing emission levels, maximizing the initial hardware investment's environmental value.  In contrast, MonetDB reaches its break-even point after only $10^{4}$ to $10^{5}$ queries, indicating less effective amortization of the manufacturing footprint due to its higher operational emissions.

To support the break-even analysis and provide insights into these break-even patterns, we measure the average power consumption during workload execution. Figure~\ref{fig:break-even-comparison}c presents the average power consumption for each database system, highlighting how power usage directly impacts environmental efficiency and break-even time. More specifically, DuckDB maintains stable power consumption across all memory configurations while achieving high carbon efficiency, indicating optimal use of its power draw to complete work. This combination enables effective amortization of manufacturing emissions. Hyper shows a gradual increase in power consumption with larger memory configurations while maintaining reasonable carbon efficiency, suggesting effective resource utilization. MonetDB exhibits moderate power consumption but lower carbon efficiency, demonstrating that low power consumption alone doesn't guarantee good environmental efficiency. StarRocks shows the highest average power consumption without proportional gains in carbon efficiency, indicating that its MPP architecture's overhead significantly impacts both immediate power consumption and long-term carbon efficiency.

The temporal perspective, presented in Figure~\ref{fig:break-even-comparison}b, reveals that all database systems achieve break-even within their first 14 months of operation, indicating that the environmental cost of manufacturing is recovered well within typical server lifespans. DuckDB's extended break-even period (200-400 days with larger memory configurations) coupled with high query amortization and stable power consumption, demonstrates an effective use of the hardware's embodied carbon, as it processes significantly more queries while accumulating fewer emissions. This efficiency becomes particularly valuable post-break-even, as continued operation yields more computational work per unit of manufacturing emissions. In contrast, MonetDB reaches break-even with fewer queries and low power consumption, indicating less efficient manufacturing footprint amortization and suggesting faster emissions accumulation during extended operation without throughput gains. Hyper shows balanced operation with memory-proportional break-even times (150-220 days) and consistent power-to-throughput ratios, indicating efficient resource scaling. StarRocks exhibits high sensitivity to memory configuration, with break-even times ranging from 100 to 200 days, but its high power consumption and lower query throughput indicate that its MPP architecture introduces overhead that impacts environmental efficiency without proportional performance gains.

These temporal patterns emphasize that database systems' environmental impact should be evaluated not only by break-even speed, but also by how efficiently they utilize hardware throughout their operational lifetime. The ability to process more queries per unit of embodied emissions becomes increasingly important as operation continues beyond the break-even point, maximizing the environmental return on the initial manufacturing investment.

\begin{figure*}[t!]
    \centering
    \begin{minipage}{0.48\textwidth} 
        \centering
        \includegraphics[width=\linewidth]{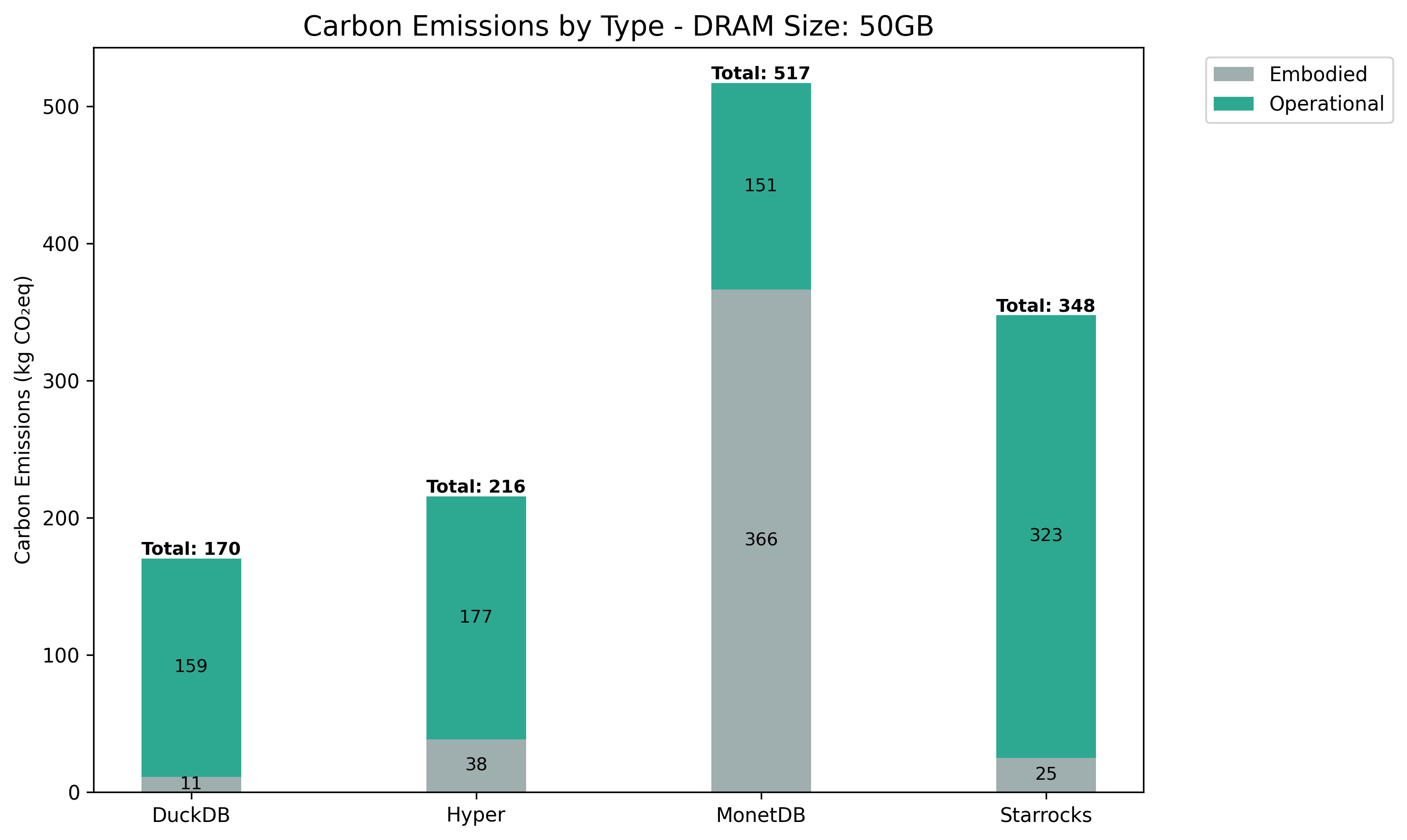}
        \caption*{(a) Total carbon emissions during the lifetime of the database server}
        \label{fig:carbon_lifetime}
    \end{minipage}
    \hspace{0.02\textwidth} 
    \begin{minipage}{0.48\textwidth} 
        \centering
        \includegraphics[width=\linewidth]{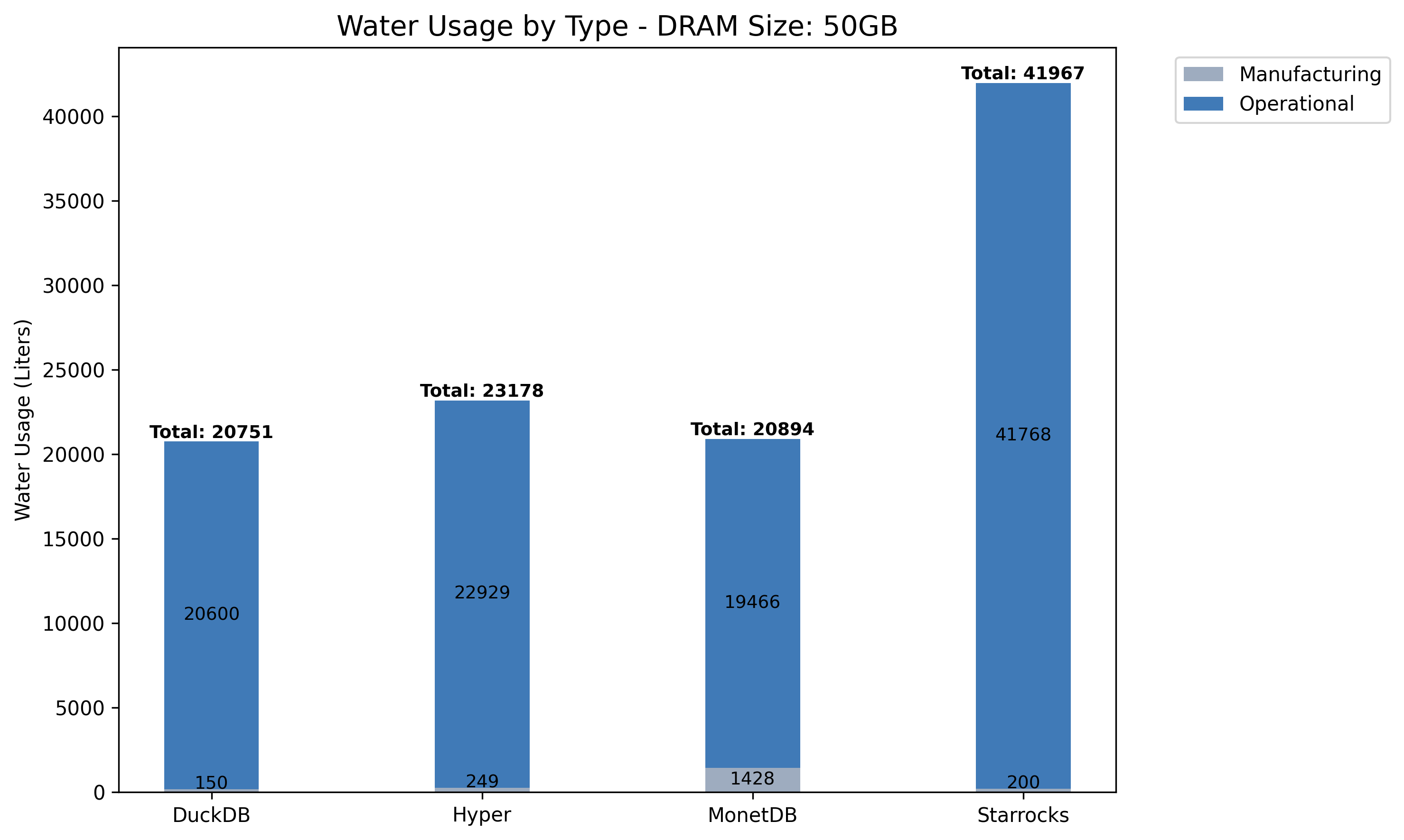}
        \caption*{(b) Total water consumption during the lifetime of the database server}
        \label{fig:water_lifetime}
    \end{minipage}
    \caption{Environmental impact metrics during server's lifetime}
    \label{fig:lifetime}
\end{figure*}

\subsubsection{\textbf{Hardware Lifespan Impact}}\hspace{-9pt}

Following ATLAS's methodology for analyzing hardware longevity, we examine how different database architectures affect storage endurance and their resulting environmental implications across a server's operational lifetime, focusing on a common but challenging scenario in database operations: processing read-intensive workloads where data sizes (100GB) exceed available memory (50GB DRAM). This memory-constrained scenario forces systems to engage in frequent disk interactions, making SSD wear analysis critical for understanding hardware replacement needs and long-term environmental costs. By examining how different architectures handle memory constraints, we can better understand the relationship between architectural choices, storage endurance, and their consequent environmental impacts.

Figure~\ref{fig:writeio_footprint}a presents write I/O footprint per TPC-H execution at scale factor 100, illustrating the amount of data written to the SSD during query processing. MonetDB exhibits significantly higher write volumes while DuckDB performs nearly negligible write operations, reflecting the fundamental architectural differences in disk I/O management. Hyper and StarRocks show moderate write volumes, suggesting more balanced approaches to storage interaction.

Projected over a year of continuous operation (Figure ~\ref{fig:writeio_footprint}b), these patterns reveal substantial cumulative wear differences. MonetDB's extensive write I/O leads to the highest annual wear, exceeding the SSD's 240TB TBW (Terabytes Written) endurance rating, necessitating multiple SSD replacements within a year. This reflects MonetDB's operator-at-a-time execution model's reliance on intermediate result generation and storage. In contrast, DuckDB's minimal write I/O footprint results in the lowest wear, indicating its efficient in-memory processing and reduced storage requirements while Hyper and StarRocks generate intermediate volumes. These variations in write patterns have direct implications for SSD longevity and replacement frequency over the server's lifetime.

The environmental impact of these storage interaction patterns becomes evident when examining the total carbon emissions and water consumption over the server's five-year lifespan (Figures~\ref{fig:lifetime}a and ~\ref{fig:lifetime}b). Our analysis incorporates two key components: operational impacts (CPU energy consumption, operational carbon emissions, and operational water footprint) and manufacturing impacts (embodied carbon emissions and manufacturing water footprint from the initial hardware, including DRAM, CPU, and SSD, and the replacement SSDs needed throughout the server's lifetime).

For carbon emissions (Figure~\ref{fig:lifetime}a), MonetDB shows the highest total at 517 kgCO2, with manufacturing emissions accounting for approximately 70\% of its total footprint. This predominantly manufacturing-driven profile reflects the environmental cost of hardware replacements necessitated by its intensive storage interaction patterns. StarRocks presents a contrasting pattern - while its total emissions are lower at 348 kgCO2, over 90\% comes from operational rather than manufacturing emissions. DuckDB achieves the lowest total carbon footprint at 170 kgCO2, maintaining a similar operational-heavy distribution to StarRocks, while Hyper maintains a moderate profile with its total of 216 kgCO2 following a comparable distribution pattern.

Water consumption (Figure~\ref{fig:lifetime}b) reveals a different hierarchy. StarRocks leads with the highest total water consumption, with operational water usage accounting for more than 99\% of its total. This presents a stark contrast to MonetDB, which despite having the highest carbon footprint, shows more modest water consumption at roughly half of StarRocks' total. Both DuckDB and Hyper maintain efficient water usage profiles at similar levels to MonetDB, with operational water usage dominating their totals at over 95\% in each case. Notably, across all systems, the manufacturing water footprint represents a much smaller proportion of total water consumption compared to the manufacturing-operational split we observed in carbon emissions.

These results highlight an interesting divergence between carbon emissions and water consumption metrics. While StarRocks exhibits the highest operational environmental metrics in both categories, its total environmental impact varies significantly between them. Its total carbon emissions remain below MonetDB's, primarily due to substantially lower manufacturing emissions. However, StarRocks leads in total water consumption, as the manufacturing water footprint contributes proportionally less to overall water usage compared to the manufacturing carbon footprint's impact on total emissions. Notably, StarRocks achieves higher query throughput than MonetDB while operating under memory constraints, illustrating the complex interplay between environmental efficiency and computational performance in analytical database systems.

This analysis reveals how architectural choices can have varying implications across different environmental metrics, particularly when operating under memory constraints where data sizes exceed available memory. The results suggest that optimizing for one environmental metric may not automatically lead to improvements in others, highlighting the importance of considering multiple environmental factors in database system design.

\subsection{Threats to Validity}
Our study has some potential threats to validity that warrant discussion. First, as shown in \cite{desrochers2016validation} and validated later in \cite{khan2018rapl}, RAPL DRAM energy measurements have accuracy limitations. While RAPL DRAM values match overall energy and power trends, they typically show a constant power offset from actual measurements.  Despite this, RAPL DRAM measurements remain useful for comparative analysis and understanding relative power consumption patterns across different workloads.

Second, carbon emission data accuracy and comparability vary across countries. As highlighted by Andrew \cite{andrew2020comparison}, variations in system boundaries, data sources, and estimation methodologies lead to significant discrepancies in reported emissions. These challenges affect emission estimates in many regions where data quality and methodological uncertainties persist. Furthermore, flawed or incomplete emission data can distort international comparisons and hinder the development of effective climate policies \cite{climate_pledges_flawed_emissions}. These discrepancies emphasize the need for caution when interpreting and comparing emission data, especially in cross-country analyses. 

\section{Conclusions}

Our comprehensive analysis of analytical database systems reveals that environmental impacts significantly transcend  traditional performance metrics. The findings demonstrate that environmental considerations encompass both operational impacts and long-term sustainability implications, with water footprint emerging as a critical metric varying significantly across architectures. The total environmental cost combines operational effects with hardware lifecycle considerations, highlighting the importance of holistic environmental assessment.

Database architecture fundamentally determines environmental efficiency through both immediate power consumption and long-term sustainability. DuckDB's embedded architecture achieves superior environmental efficiency while maintaining performance, while the contrast between MonetDB's operator-at-a-time approach and DuckDB's vectorized execution demonstrates the environmental implications of execution model choice. Also, environmental efficiency encompasses the interplay of hardware utilization, execution patterns, and energy usage, with database systems typically amortizing manufacturing carbon footprint within 14 months, though timing varies by architecture.

Perhaps most significantly, our research establishes that geographical location can have a greater impact on environmental footprint than database choice. The carbon intensity and water footprint of local power grids dramatically affect operational emissions, sometimes exceeding the impact of architectural decisions. This finding suggests that deployment location should be a key consideration in environmental strategy, particularly for distributed database systems, where data and processing are spread across multiple geographical locations.

Looking ahead, several critical areas warrant further investigation. First, understanding how query optimization decisions and execution plans impact the environmental footprint could reveal opportunities for "green" query planning that considers both performance and environmental costs. Second, analyzing the relationship between low-level performance counters and environmental impact could help identify specific hardware-level bottlenecks that contribute to increased energy consumption and emissions. Third, investigating how concurrent query execution affects environmental efficiency could provide insights into optimizing workload scheduling for reduced environmental impact. Additionally, future research should focus on developing database architectures that optimize for both performance and environmental efficiency, including adaptive systems that can leverage clean energy availability and metrics that better capture the full environmental impact of database operations.

\section*{Acknowledgements}
This work has in part been supported by National Sciences and Engineering Research Council
of Canada (NSERC) and Ontario Research Fund (ORF).

\bibliographystyle{unsrt}
\bibliography{sample}

\end{document}